\documentclass[a4paper,11pt]{article}

\usepackage{jheppub}
\usepackage{xcolor,colortbl}
\allowdisplaybreaks

\def\MSbar{\overline{\mathrm{MS}}}

\def\ep{\epsilon}
\def\z#1{{\zeta_{#1}}}
\def\ca{{C^{}_A}}
\def\cf{{C^{}_F}}
\def\tf{{T^{}_F}}
\def\nf{{n^{}_f}}
\def\nl{{n^{}_l}}
\def\nh{{n^{}_h}}

\def\as{{\alpha_s}}
\def\asb{{\alpha_s^0}}
\def\asnf{{\alpha_s^{(\nf)}}}
\def\asnl{{\alpha_s^{(\nl)}}}

\def\lmu{l_\mu}
\def\lmu#1{l^{#1}_\mu}

\def\lb#1{\if 1#1 \ln\beta \else \ln^#1\beta \fi}
\def\lt#1{\if 1#1 \ln 2 \else \ln^#1 2 \fi}

\title{Virtual amplitudes and threshold behaviour of hadronic
  top-quark pair-production cross sections}

\author{P. B\"arnreuther,}
\author{M. Czakon}
\author{and P. Fiedler}
\affiliation{Institute for Theoretical Particle Physics and Cosmology, 
RWTH Aachen University, D-52056 Aachen, Germany}

\emailAdd{pbaernreuther@physik.rwth-aachen.de}
\emailAdd{mczakon@physik.rwth-aachen.de}
\emailAdd{fiedler@physik.rwth-aachen.de}

\abstract{We present the two-loop virtual amplitudes for the
  production of a top-quark pair in gluon fusion. The evaluation
  method is based on a numerical solution of differential equations
  for master integrals in function of the quark velocity and
  scattering angle starting from a boundary at high-energy. The
  results are given for the renormalized infrared finite remainders on
  a large grid and have recently been used in the calculation of the
  total cross sections at the next-to-next-to-leading order. For
  convenience, we also give the known results for the quark
  annihilation case on the same grid. Outside of the kinematical range
  covered by the grid, we provide threshold and high-energy
  expansions.

  From expansions of the two-loop virtual amplitudes, we determine the
  threshold behavior of the total cross sections at
  next-to-next-to-leading order for the quark annihilation and gluon
  fusion channels including previously unknown constant terms. In our
  analysis of the quark annihilation channel, we uncover the presence
  of a velocity enhanced logarithm of Coulombic origin, which was
  missed in a previous study.
}

\dedicated{\rm TTK-13-28}

\begin{document} 
\maketitle
\flushbottom


\section{Introduction}

Results for next-to-next-to-leading order corrections to hadronic cross
sections for top-quark pair \cite{Baernreuther:2012ws, Czakon:2012zr,
  Czakon:2012pz, Czakon:2013goa}, di-jet \cite{Ridder:2013mf,
  Currie:2013dwa}, and Higgs + jet \cite{Boughezal:2013uia} production
demonstrate the recently achieved tremendous progress in perturbative
techniques. On the one hand, these developments were made possible by
research on subtraction schemes, in particular antenna subtraction
\cite{GehrmannDeRidder:2005cm, Currie:2013vh} and sector improved
residue subtraction, {\tt STRIPPER}, \cite{Czakon:2010td,
  Czakon:2011ve}. On the other hand, the case of top-quarks required
the application of semi-numerical methods \cite{Czakon:2008zk} for the
evaluation of two-loop virtual amplitudes. It is the purpose of the
present publication to present the results for these amplitudes in the
gluon fusion channel thus completing the documentation of the ingredients
necessary to reproduce the cross section results of
Ref.~\cite{Czakon:2013goa}.

Beyond the one-loop level, most of the known virtual amplitudes have
been obtained by a reduction of the Feynman integrals using
integration-by-parts identities \cite{Chetyrkin:1981qh} and the
Laporta algorithm \cite{Laporta:2001dd}, followed by the evaluation of
the occurring master integrals, either by Mellin-Barnes methods
\cite{Smirnov:1999gc, Tausk:1999vh} or analytic solution of
differential equations in the kinematic invariants
\cite{Kotikov:1990kg, Remiddi:1997ny}. As far as top-quark
pair-production amplitudes are concerned, there is an on-going effort
along these lines, aiming at fully analytic results
\cite{Bonciani:2008az, Bonciani:2009nb, Bonciani:2010mn,
  vonManteuffel:2013uoa, Bonciani:2013ywa}. While differential
equations provide an iterative algorithm for the evaluation of the
integrals, the main problem is to find an appropriate basis of special
functions to express the results. An alternative strategy has been
proposed in Ref.~\cite{Czakon:2008zk} and applied to the
quark-annihilation channel. Instead of solving the differential
equations analytically, the idea was to resort to numerical
methods. The problems of this approach are of two kinds. At first, it
is necessary to provide a boundary condition in the form of high
precision values of the integrals at a single point. Inspired by
Refs.~\cite{Czakon:2007ej, Czakon:2007wk}, a point in the high-energy
range has been chosen for this purpose. The asymptotics of the
integrals in this limit have been derived using Mellin-Barnes
techniques. The second problem is related to singularities of the
differential equations, which cause substantial problems. In the case
of the quark-annihilation channel amplitudes the use of higher
numerical precision was sufficient to provide numerical values within
some acceptable kinematical domain. Unfortunately, the gluon fusion
channel is substantially more demanding both in the determination of
the asymptotics of the integrals, and in the treatment of numerical
instabilities. The techniques that we have applied in this case are
documented in this work.

The presentation of numerical results is always a non-trivial
issue. If the functional dependence of the amplitudes is smooth
enough, one might consider giving a grid of values. This is indeed
what we have done. One must, however, remember that due to well-known
singularities (mostly Coulomb and collinear), the amplitudes diverge
in some corners of the phase space. A result in the form of a grid is
not practicable there. Instead, we provide expansions beyond the
borders of the grid, both close to the threshold and far away from
it. Finally, even though the quark-annihilation amplitudes are known
since Ref.~\cite{Czakon:2008zk}, we reproduce them here in the
same format as those of the gluon-fusion channel.

The two-loop amplitudes presented here have already been used for the
calculation of total cross sections. Moreover, they will be used
in the near future for the evaluation of differential distributions,
and will certainly serve as a benchmark for on-going analytic
calculations. There is yet another application of our results, which
concerns the threshold expansion of the total cross sections. In
Ref.~\cite{Czakon:2013hxa}, we have discussed how to derive such an
expansion from soft-gluon factorization using virtual amplitudes and
soft functions. In this work, we will apply these methods and derive
the leading threshold effects including quark-velocity independent terms. This
is an extension of the results of
Ref.~\cite{Beneke:2009ye}. Interestingly, while the study presented in
the latter reference aimed at the derivation of all the terms singular
in the velocity of the top-quark, our new result shows a discrepancy
in the form of a logarithm of the velocity, which is due to Coulombic
effects. We will elucidate the origin of the omission.

The paper is organized as follows. In the next section, we introduce
our notation, and discuss ultraviolet and infrared renormalization of
the amplitudes. Subsequently, we describe the evaluation of the
high-energy asymptotics of the master integrals and the
numerical/semi-numerical solution of differential equations for
arbitrary kinematics including expansions at singular points. We
continue with the presentation of the results for the amplitudes
including benchmark numerical values, and threshold
expansion. Finally, we present the threshold expansions of total cross
sections. The main text is closed with conclusions and followed by
appendices containing renormalization constants and anomalous
dimensions, as well as cross section expansions for arbitrary
color. The  numerical results for the amplitudes on a phase space
grid, and their high energy expansions are attached to the electronic
submission of this paper.


\section{Preliminaries}

We consider the virtual amplitudes for two processes: the gluon-fusion
channel heavy-quark pair-production
\begin{equation}
\label{eq:ggQQ}
g(p_1) + g(p_2) \:\:\rightarrow\:\: Q(p_3,m) + {\bar Q}(p_4,m) \, ,
\end{equation}
and the quark-annihilation channel heavy-quark pair-production
\begin{equation} q(p_1) + {\bar q}(p_2) ~ \rightarrow ~ Q(p_3,m) + {\bar
Q}(p_4,m) \, .
\end{equation}
In both cases, the final state heavy-quarks (top-quarks) are on-shell,
$p_3^2 = p_4^2 = m^2$. The results presented in the next sections for
the first process are new, whereas those for the second process are an
improvement, as far as the kinematical range is concerned, over those
given in Ref.~\cite{Czakon:2008zk}. The amplitudes can be expressed in
terms of modified Mandelstam invariants
\begin{equation}
s = (p_1+p_2)^2 , ~ t  =  m^2 - (p_1-p_3)^2 , ~ u  = m^2 - (p_1-p_4)^2 ,
\end{equation}
which satisfy $s = t+u$, where
\begin{equation}
t = \frac{s}{2}(1 - \beta \cos \theta) \; ,
\end{equation}
and $\theta$ is the scattering angle. The ratio of the mass and the
center-of-mass energy is expressed in terms of the velocity of the
top-quark
\begin{equation}
\beta = \sqrt{1-\frac{4m^2}{s}} .
\end{equation}
This is a natural variable for current hadron collider applications,
where it is known that the bulk of the top-quark pair-production cross
section comes from the range of moderate velocities. The high-energy
limit is condensed to $\beta \approx 1$, while the production
threshold is to be found at $\beta \approx 0$. 

The bare on-shell amplitudes admit a perturbative expansion, of which
only the first three terms are relevant
\begin{eqnarray}
  | {\cal M}^0_{g,q}(\asb,m^0,\ep) \rangle
  & = &
  4 \pi \asb \left[
  | {\cal M}_{g,q}^{(0)}(m^0,\ep) \rangle
  + \left( {\asb \over 2 \pi} \right) | {\cal M}_{g,q}^{(1)}(m^0,\ep) \rangle
  + \left( {\asb \over 2 \pi} \right)^2 | {\cal M}_{g,q}^{(2)}(m^0,\ep) \rangle
  \right] . \nonumber \\
\end{eqnarray}
The subscript $g$ or $q$ specifies the initial state, while the
ket-notation signifies a vector in color and spin space. We have also
indicated the dependence on the parameter of dimensional
regularization with $d = 4-2\ep$ space-time dimensions. The
external degrees of freedom are treated with Conventional Dimensional
Regularization (CDR), in particular the gluons have $d-2$ degrees of
freedom. As usual, we are interested in color and spin summed squared
amplitudes. The two-loop contributions are then given as a product of
the Born and two-loop amplitudes, and may be written as an
expansion in the number of colors, $N_c$, of a general
$\mbox{SU}(N_c)$ group, and the number of closed light-, $\nl$, and
heavy-quark, $\nh$, loops. The heavy quarks in the loops are assumed
to have the same mass as the external top-quarks. We write
\begin{eqnarray}
\label{eq:colordecgg}
 2 {\rm Re}\, \langle {\cal M}_g^{(0)} | {\cal M}_g^{(2)} \rangle &=&
(N_c^2-1) \nonumber \\ &\times& \biggl(
N_c^3 A^{(g)} + N_c B^{(g)}  + {1 \over N_c} C^{(g)} + {1 \over N_c^3} D^{(g)}
+ N_c^2 \nl E_l^{(g)} + N_c^2 \nh E_h^{(g)} \nonumber \\ 
& &
+ \nl F_l^{(g)} + \nh F_h^{(g)}
+ {\nl \over N_c^2} G_l^{(g)} + {\nh \over N_c^2} G_h^{(g)}
+ N_c \nl^2 H_l^{(g)} + N_c \nl \nh\, H_{lh}^{(g)} \nonumber \\
& & 
+ N_c \nh^2 H_h^{(g)}
+ {\nl^2 \over N_c} I_l^{(g)} + {\nl \nh \over N_c} I_{lh}^{(g)} +
{\nh^2 \over N_c} I_h^{(g)}
\biggr)
\, , \\ \nonumber \\
\label{eq:colordecqq}
2 {\rm Re}\, \langle {\cal M}_q^{(0)} | {\cal M}_q^{(2)} \rangle &=& 2
(N_c^2-1) \nonumber \\ &\times& \biggl(
N_c^2 A^{(q)} + B^{(q)} + {1 \over N_c^2} C^{(q)}
+ N_c \nl D_l^{(q)} + N_c \nh D_h^{(q)}
+ {\nl \over N_c} E_l^{(q)} + {\nh \over N_c} E_h^{(q)} \nonumber \\
& & + \nl^2 F_l^{(q)} + \nl \nh F_{lh}^{(q)} + \nh^2 F_h^{(q)}
\biggr) .
\end{eqnarray}
The factor $2$ difference between the prefactors in the equations has
no deeper meaning, but is taken over from
\cite{Czakon:2007ej, Czakon:2007wk}.

The amplitudes are renormalized according to
\begin{equation}
\left| {\cal M}^R_{g,q}\left(\asnf(\mu),m,\mu,\ep\right)
\right\rangle = \left( \frac{\mu^2 e^{\gamma_E}}{4\pi}\right)^{-2\ep}
Z_{g,q} Z_{Q} | {\cal M}_{g,q}(\asb,m^0,\ep) \rangle \; ,
\end{equation}
where $Z_g,Z_q$ and $Z_Q$ are the on-shell wave-function renormalization
constants for gluons, and light- and heavy-quarks respectively. A
prefactor has been introduced in order to keep the amplitudes
dimensionless and avoid unnecessary $\gamma_E - \ln(4\pi)$ terms. The
bare mass is just $m^0 = Z_m m$, while the bare coupling constant is
\begin{eqnarray}
\asb = \left( \frac{e^{\gamma_E}}{4\pi}\right)^\ep
\mu^{2\ep} Z^{(\nf)}_\as \asnf(\mu) \; ,
\end{eqnarray}
which corresponds to the $\MSbar$ scheme with $n_f = n_l
+ n_h$ active flavors, if the loop integrals are calculated with the
measure $d^d k/(2\pi)^d$. The necessary renormalization constants can
be found in Appendix~\ref{sec:RConst}. Notice that our results are
always given with $\mu = m$.

Renormalization with $\nl+\nh$ active flavors has been assumed in
every previous calculation of two-loop virtual amplitudes for
heavy-quark pair-production. However, once we are interested in a
hadronic cross section, we have to reassess the question of the number
of active flavors. Indeed a natural scale for top-quark
pair-production is set by the top-quark mass. The bulk of the cross
section comes from the regime, where non-logarithmic top-quark mass
effects are not negligible, since the top-quarks themselves are not
ultra-relativistic. It is thus more appropriate to calculate cross
sections with $n_l$ active flavors and not resum the mass logarithms
in either the coupling constant or the parton distribution
functions. Amplitudes with $n_l$ active flavors are obtained by
decoupling in $\as$, which amounts to the substitution
\begin{equation}
\asnf = \zeta_\as \asnl \; ,
\end{equation}
where the decoupling constant, $\zeta_\as$, has a perturbative
expansion in the strong coupling constant and can be found in the
Appendix~\ref{sec:RConst}.

Although the ultraviolet renormalized amplitudes still contain divergences of
infrared origin, the structure of these divergences is completely
understood at the two-loop level \cite{Aybat:2006mz, Mitov:2009sv,
  Becher:2009qa, Becher:2009kw, Czakon:2009zw, Ferroglia:2009ii,
  Mitov:2010xw}. We, therefore, define the finite remainder of the
amplitudes through infrared renormalization
\begin{equation}
\left| {\cal M}^{fin}_{g,q} \left(\asnl,m,\mu
\right) \right\rangle =
{\bf Z}_{{\cal M}_{g,q}}^{-1}(\{\underline{p}\},\{\underline{m}\},\mu,\ep)
\left| {\cal M}^R_{g,q} \left(\asnl,m,\mu,\ep \right) \right\rangle
\; ,
\end{equation} 
where we have underlined that the amplitude on the right hand side of
the equation has been obtained by decoupling of the heavy quark from
the running of $\as$.
The $\MSbar$ renormalization constant ${\bf Z}_{{\cal M}_{g,q}}$ is
a matrix in color space and has a non-trivial dependence on the 
kinematics $\{\underline{p}\}=\{p_1,...,p_4\}$, and by the same on the masses
$\{\underline{m}\}=\{0,0,m,m\}$ of the external partons. It
can be derived from the differential equation
\begin{equation}
\frac{d}{d\ln\mu}\,{\bf
  Z}_{\cal M}(\{\underline{p}\},\{\underline{m}\},\mu,\ep)\,=\,-{\bf
  \Gamma}_{\cal M}(\{\underline{p}\},\{\underline{m}\},\mu) \,
{\bf Z}_{\cal M}(\{\underline{p}\},\{\underline{m}\},\mu,\ep) \; ,
\end{equation} 
where we do not specify the initial state anymore, and the color space
matrix anomalous dimension is given by up to two-loops by
\cite{Ferroglia:2009ii}
\begin{equation}\label{eq:IRGamma}
\begin{split}
{\bf \Gamma}_{\cal M}(\{\underline{p}\},\{\underline{m}\},\mu)\,
& = \,\sum\limits_{(i,j)}\frac{{\bf T}_i\cdot{\bf
    T}_j}{2}\,\gamma_{\text{cusp}}\left(\asnl\right)\,
\ln\frac{\mu^2}{-s_{ij}}\,+\,\sum\limits_i
\gamma^i\left(\asnl\right)\\ 
& \hspace{-2.5cm} -\,\sum\limits_{(I,J)}\frac{{\bf T}_I\cdot{\bf T}_J}{2}\,
\gamma_{\text{cusp}}\left(\beta_{IJ},\asnl\right)\,+\,\sum\limits_I
\gamma^I\left(\asnl\right)\,
+\,\sum\limits_{I,j}{\bf T}_I\cdot{\bf
  T}_j\,\gamma_{\text{cusp}}\left(\asnl\right)\,
\ln\frac{m_I\,\mu}{-s_{Ij}}\\
& \hspace{-2.5cm} +\,\sum\limits_{(I,J,K)}i\,f^{abc}\,{\bf T}_I^a\,{\bf T}_J^b\,
{\bf T}_K^c\,F_1(\beta_{IJ},\beta_{JK},\beta_{KI})\\ 
& \hspace{-2.5cm} +\,\sum\limits_{(I,J)}\sum\limits_k\,i\,f^{abc}\,
{\bf T}_I^a\,{\bf T}_J^b\,{\bf T}_k^c\,f_2\left(\beta_{IJ},
\ln\frac{-\sigma_{Jk}\,v_J\cdot p_k}{-\sigma_{Ik}\,
v_I\cdot p_k}\right)
\; .
\end{split}
\end{equation} 
The summations run over massless (indices $i, j,
k$) and massive (indices $I, J, K$) partons, with the notation $(i,j,...)$
denoting unordered tuples of different indices. The color operators
${\bf T}^a_i$ act on the color indices of the respective
partons.  If the particle is a gluon carrying a color index $c$, we
have $({\bf T}^a)_{bc}=-i\, f^{abc}$, assuming the result has been
projected on color index $b$. Similarly, for an outgoing
quark (or incoming anti-quark) the generator is $({\bf T}^a)_{bc} =
T^a_{bc}$, whereas for an incoming quark (or outgoing anti-quark) the
generator is $({\bf T}^a)_{bc} = -T^a_{cb}$. The kinematic dependence
is contained in $s_{ij} = 2\sigma_{ij} p_i \cdot p_j + i0^+$, where
the sign factor $\sigma_{ij} = +1$ if the momenta $p_i$ and $p_j$ are
both incoming or outgoing, and $\sigma_{ij} = -1$ otherwise. For
massive partons there is $p_I^2 = m_I^2$, $v_I = p_I/m_I$, and $\cosh
\beta_{IJ} = -s_{IJ}/2m_I m_J$. It was noted in Refs.~\cite{Czakon:2009zw,
  Ferroglia:2009ii} that the triple-color-correlators in the third and
fourth line of the anomalous dimension do not contribute to top
quark pair production amplitudes' divergences. The general case was
analyzed in Ref.~\cite{Czakon:2013hxa} with the conclusion that these
terms never contribute to color and spin summed amplitudes, as long as
the latter are real. The anomalous dimensions' coefficients relevant
to the present work are given in Appendix~\ref{sec:RConst}.

With the two definitions of ultraviolet and infrared (finite
remainder) renormalized amplitudes, we are ready to proceed with the
description of the computational methods and results.


\section{Virtual amplitudes}

\subsection{Methods}

The two-loop amplitudes for heavy-quark pair-production are expressed
through 726 Feynman diagrams in the gluon-fusion channel, and 190
diagrams in the quark annihilation channel. Due to the structure of
the QCD vertices, the topologies present in the quark-annihilation
case are a subset of those present in the gluon case. Using the
Laporta algorithm, the occurring integrals are reduced to a set of 422
masters, out of which only 145 are needed for the quark-annihilation
case. Based on experience, we consider integrals with less than six
propagators as easy. This leaves 212 difficult six and seven line
integrals to evaluate. In our work, we do not use any external input,
i.e.\ we do not rely on integrals calculated by others. Nevertheless,
38 of the difficult integrals have been evaluated by one of us in
Ref.~\cite{Czakon:2008zk}. The remaining number is further reduced by
the fact that 89 integrals can be obtained from others by a $t
\leftrightarrow u$ transformation. Thus, the final number of new
integrals evaluated in this work is 100. Thanks to the work
Ref.~\cite{Czakon:2007wk}, 17 of these were at least known in the
high-energy limit.

As explained in the introduction, we chose to work as proposed in
Ref.~\cite{Czakon:2007wk} and solve differential equations for the
master integrals with numerical methods starting from a boundary at
high-energy. Since we are dealing with four-point amplitudes with a
single mass scale, the integrals, once stripped of their global
mass dimension by appropriate rescaling, depend on two dimensionless
variables. This means that the functional dependence of the integrals
is fully specified by two systems of homogeneous linear first order
partial differential equations
\begin{eqnarray}
m^2 \frac{\partial}{\partial m^2} M_i(m^2/s, \, t/s, \, \ep) &=&
\sum_j J^{(m^2)}_{ij}(m^2/s, \, t/s, \, \ep) \, M_j(m^2/s, \, t/s, \,
\ep) \; ,
\nonumber \\
t \frac{\partial}{\partial t} M_i(m^2/s, \, t/s, \, \ep) &=&
\sum_j J^{(t)}_{ij}(m^2/s, \, t/s, \, \ep) \, M_j(m^2/s, \, t/s, \, \ep) \; ,
\end{eqnarray}
which can be obtained by taking derivatives in the parameters and
reducing the resulting integrals with integration-by-parts identities
to the original masters. The elements of the Jacobi matrices,
$J^{(m^2)}$ and $J^{(t)}$, are rational functions of $m^2/s$, $t/s$ and
$\ep$. We require a solution for the master integrals in the form of
Laurent expansions in $\ep$. The original equations are, therefore,
transformed into
\begin{eqnarray}
\label{eq:diffeq}
m^2 \frac{\partial}{\partial m^2} \tilde{M}_i(m^2/s, \, t/s) &=&
\sum_j \tilde{J}^{(m^2)}_{ij}(m^2/s, \, t/s) \, \tilde{M}_j(m^2/s, \,
t/s) \; ,
\nonumber \\
t \frac{\partial}{\partial t} \tilde{M}_i(m^2/s, \, t/s) &=&
\sum_j \tilde{J}^{(t)}_{ij}(m^2/s, \, t/s) \, \tilde{M}_j(m^2/s, \, t/s) \; ,
\end{eqnarray}
by expanding all quantities in $\ep$ to sufficient order. The tilde
denotes the coefficients of the respective expansions.

The solution of the system Eq.~(\ref{eq:diffeq}) is obtained by choosing
a path, possibly complex, in the parameter space
\begin{equation}
(m^2/s, \, t/s) \, \rightarrow \, (m^2(z)/s, \, t(z)/s) \; ,
\end{equation}
and solving the resulting ordinary differential equation in $z$. This
can be done either numerically or as a power-logarithmic series in
$z$. In practice, we have proceeded as follows
\begin{enumerate}

\item We have determined analytically the first few terms of the
  high-energy expansion of the master integrals. The results are a
  power-logarithmic series in $m^2/s$, with coefficients, which are exact in
  $t/s$. In order to obtain our results, we have used Mellin-Barnes
  techniques, and in particular relied heavily on the \textsc{MB}
  package \cite{Czakon:2005rk}. In a few cases, we recovered the
  exact dependence on $t$ starting from the limiting behavior at $t = 0$
  and using differential equations in $t$. In order to obtain the
  boundary condition, we performed a double expansion
  of the Mellin-Barnes representation of a given Feynman integral in
  $m^2$ and $t$. The resulting Mellin-Barnes integrals, which were
  pure numbers were evaluated with very high precision and resummed
  with the \textsc{PSLQ} algorithm \cite{pslq}. In simpler cases, we
  have used the \textsc{XSummer} package \cite{Moch:2005uc} for
  resummation.

\item In a next step, we have obtained deep high-energy expansions
  using the differential equations in $m^2$ and the boundary
  conditions from the previous step. These expansions were used to
  derive high-energy expansions of the amplitudes, and to obtain high
  precision numerical values of the master integrals at small mass.

\item Using the numerical values determined in the previous step, we
  have solved the differential equations in $m^2$ and $t$ along
  contours in the complex plane. To obtain the solution, we have used
  the software from Ref.~\cite{vode} with improvements to handle
  higher precision numbers \cite{qd}.

\item Around $\beta = 0$, we have obtained, with the help of the
  differential equations, deep power-logarithmic expansions in $\beta$
  for fixed values of the scattering angle. These expansions were
  generated from unknown boundary coefficients, which were determined
  by matching the expansion to the numerical solution from the
  previous step at a point, at which both the expansion and the
  numerical solution provide high precision. This method can be used
  at any singular point, and allows to avoid numerical instabilities
  of the differential equations.

\end{enumerate}

\subsection{Results}

The results obtained with the methods of the previous subsection fall
into three kinematical domains: threshold, ``bulk'', and high-energy. The
``bulk'' domain covers moderate $\beta$ values and is given purely
numerically on a large grid. The sampling values of $\beta$ are chosen as in
\cite{Baernreuther:2012ws, Czakon:2012zr, Czakon:2012pz,
  Czakon:2013goa}, i.e.
\begin{equation}
\beta_i = i/80 \; , \;\;\;\; i = 1, ..., 79 \; ,
\end{equation}
and $\beta_{80} = 0.999$. This covers the range of values available at
LHC @ 8 TeV. The dependence on the scattering angle is described
through
\begin{equation}
\cos\theta_i = \pm x_i \; , \;\;\;\; i = 1, ..., 21 \; ,
\end{equation}
where the $x_i$ correspond to the 21 sampling points of the Gauss-Kronrod
integration rule of order 10, and can be obtained with any major
algebraic/numeric computer system, e.g.\ \textsc{Mathematica}. The
Gauss-Kronrod rule is an efficient deterministic rule for smooth functions, which also
provides an error estimate by sampling every second point of the rule
with appropriate weights. This specific choice of the $\cos\theta$
points has been made, because a first aim was to provide very precise
values of the contributions of the amplitudes to the total cross
sections.

\begin{figure}[h]
\begin{center}
\begin{tabular}{ccc}
\includegraphics[width=7cm,angle=0]{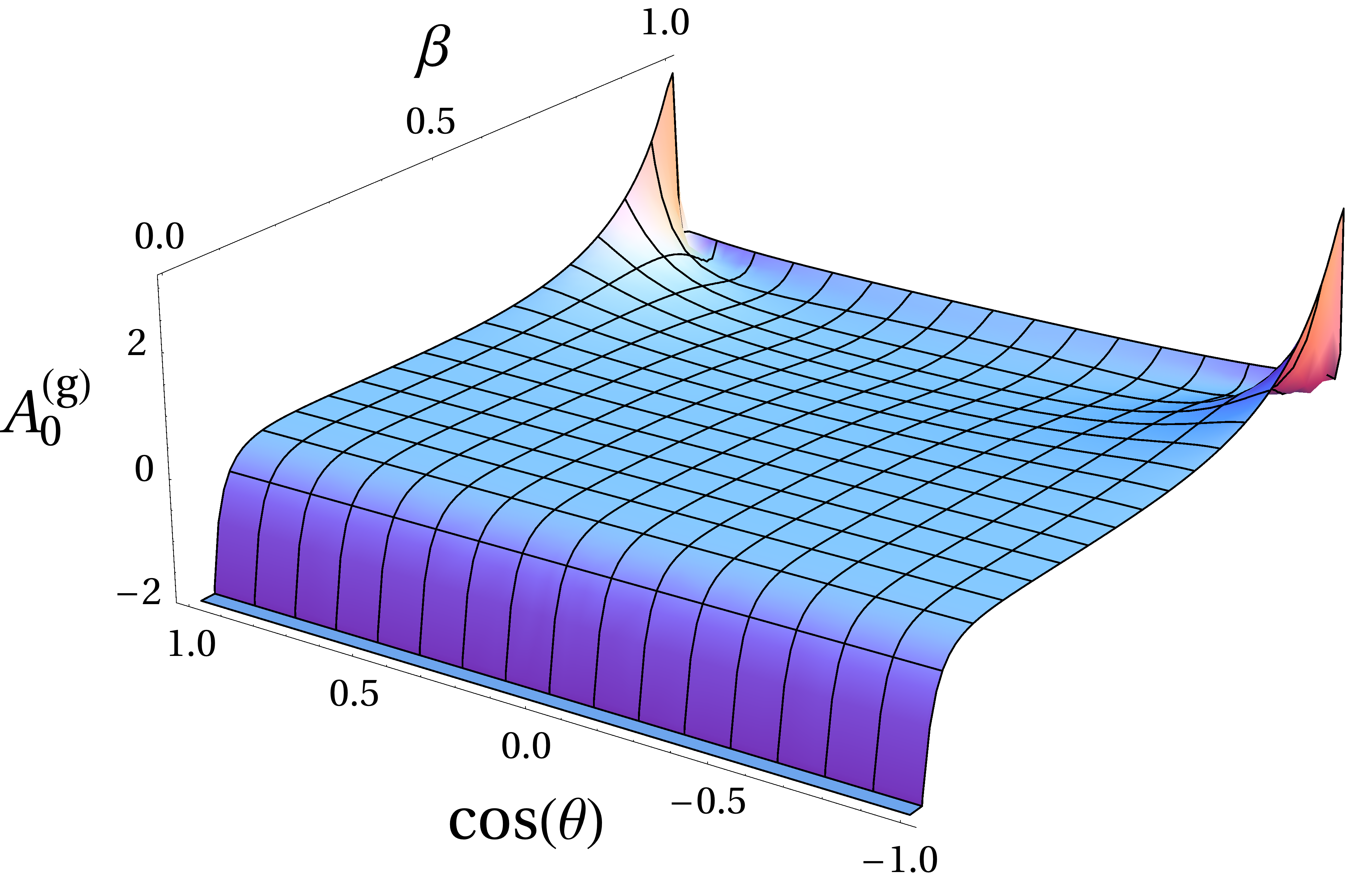}&      &\includegraphics[width=7cm,angle=0]{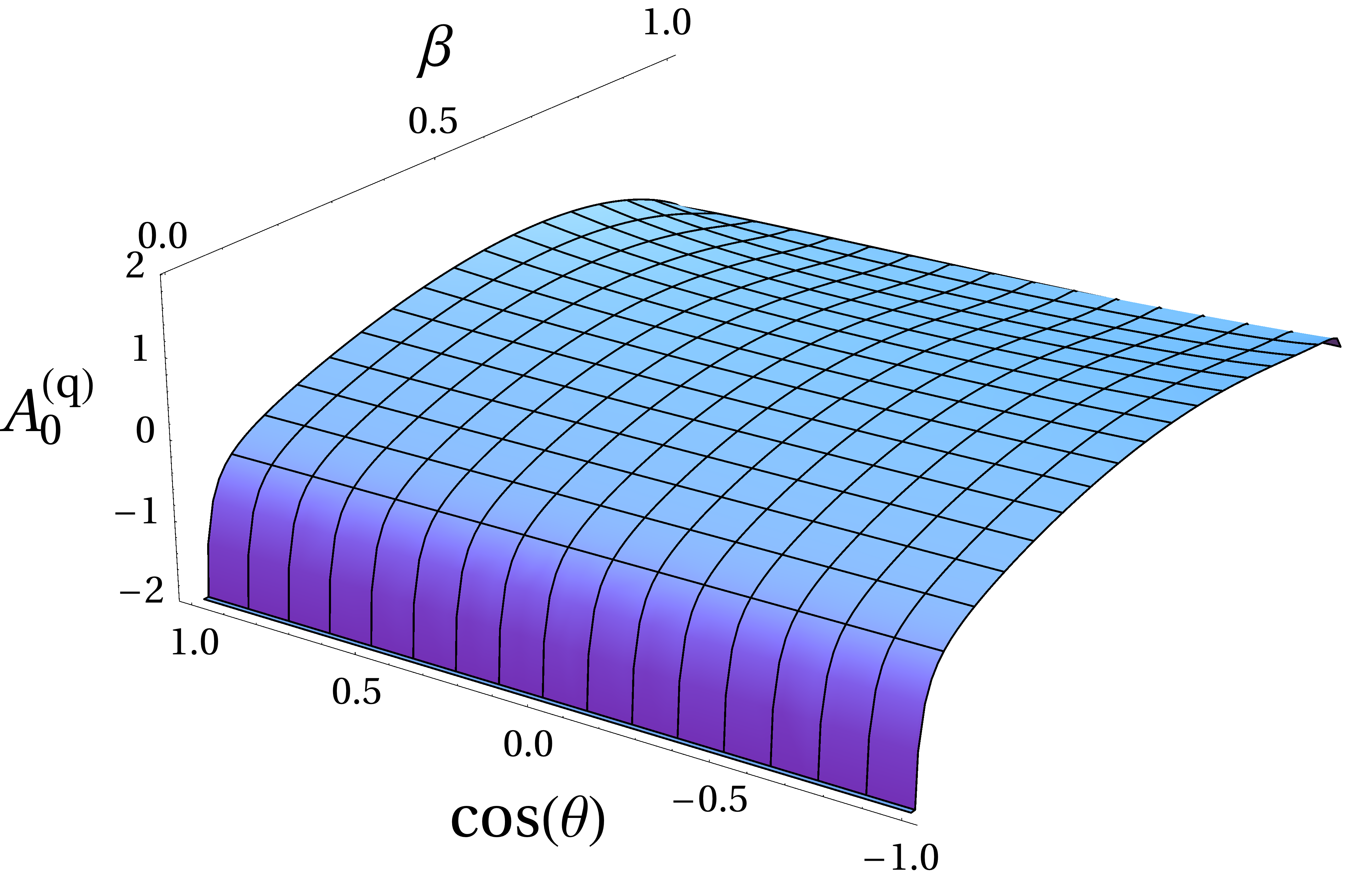}\\
\includegraphics[width=7cm,angle=0]{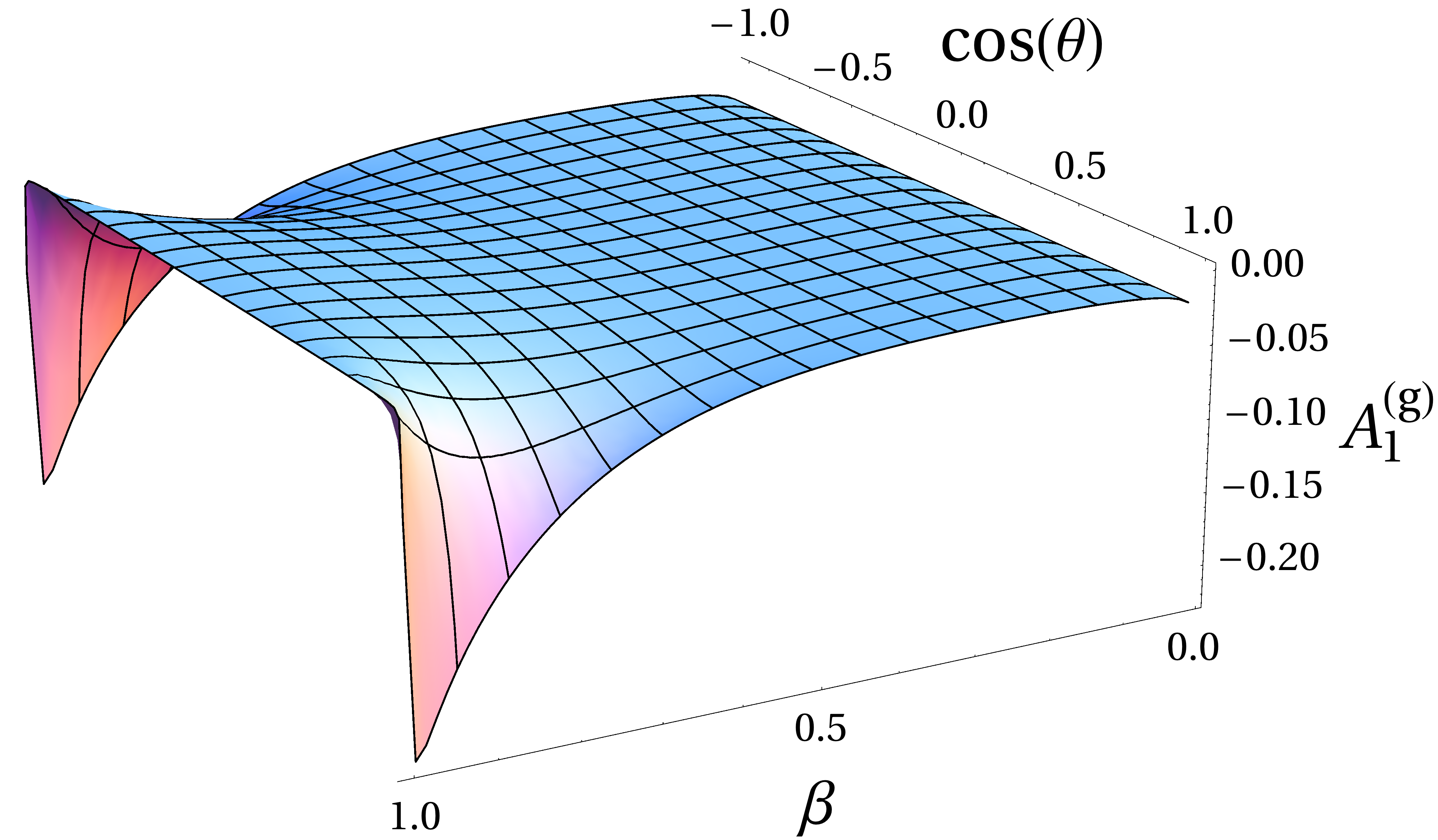}&      &\includegraphics[width=7cm,angle=0]{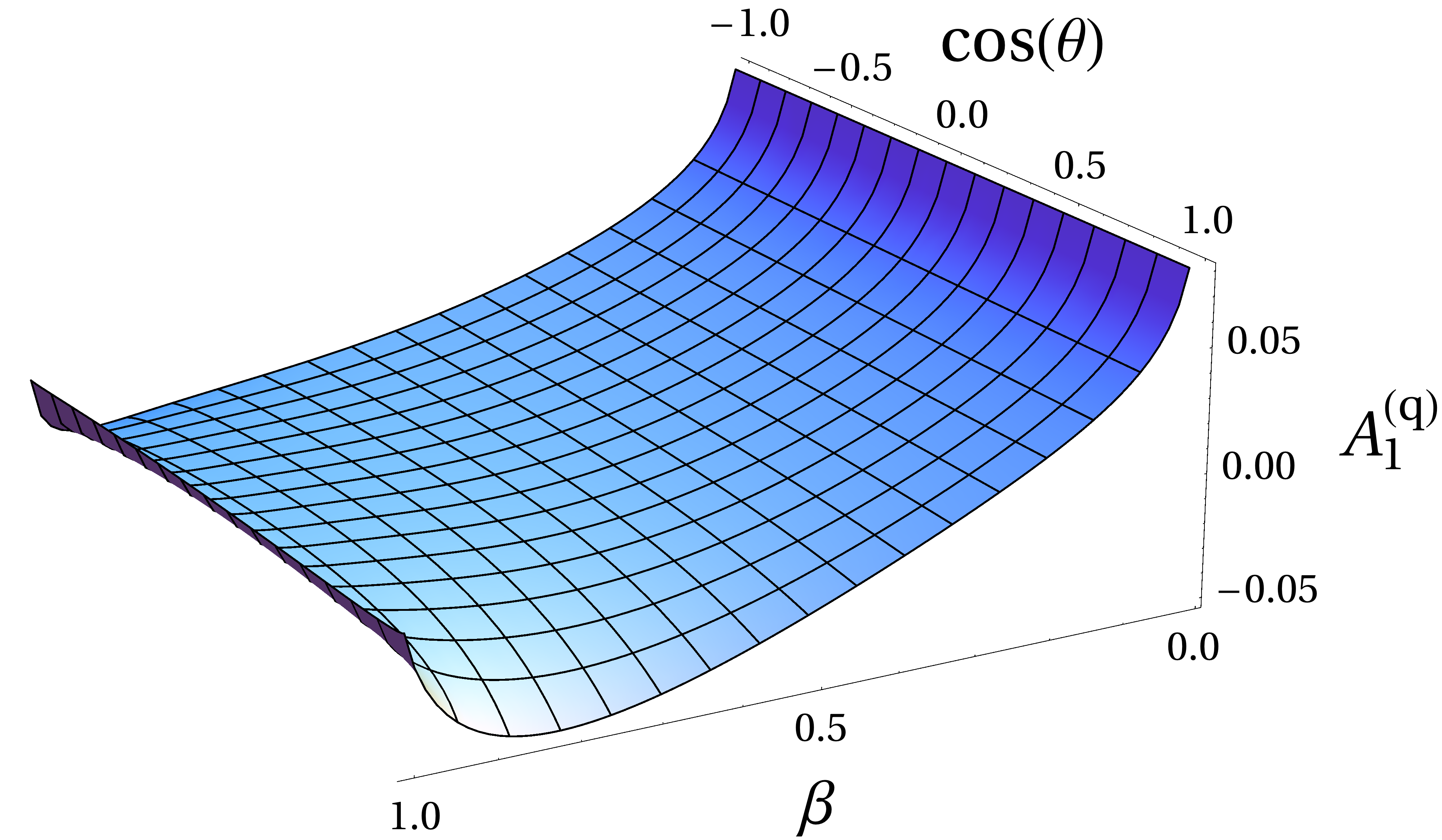}\\
\includegraphics[width=7cm,angle=0]{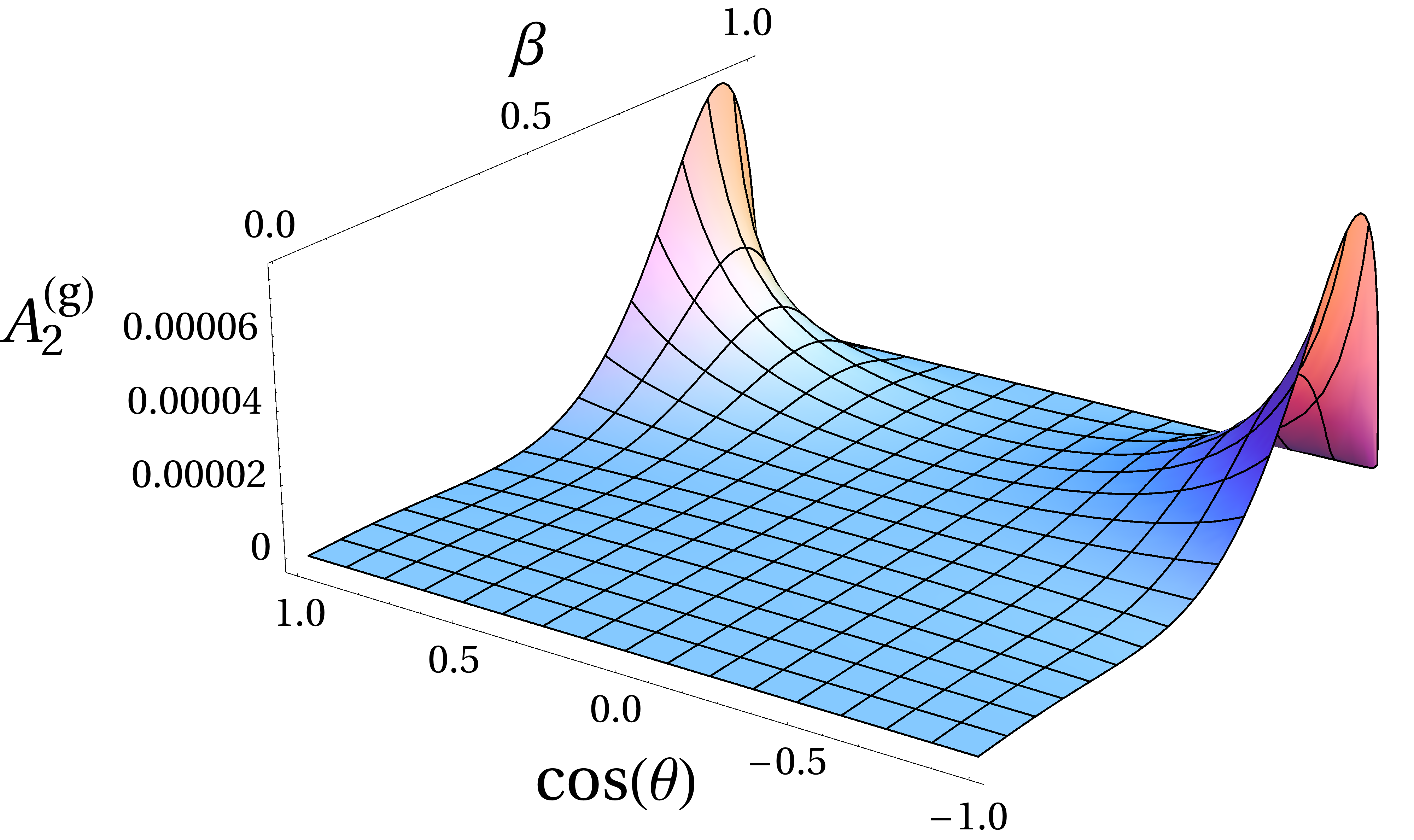}&      &\includegraphics[width=7cm,angle=0]{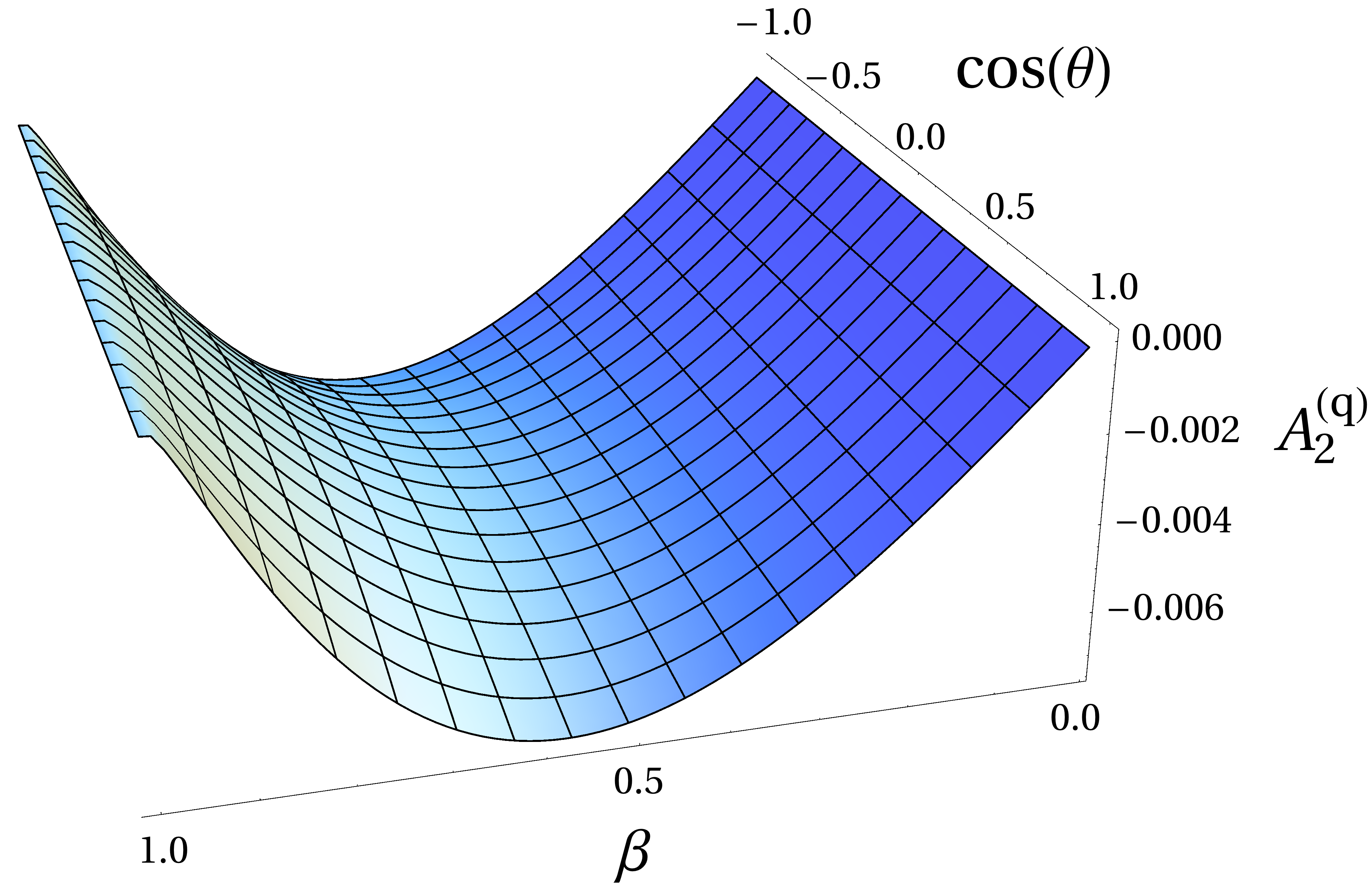}\\
\end{tabular}
\end{center}
\caption{\label{fig:amplitudes} \sf Finite remainders of the
  gluon-fusion (left) and quark-annihilation (right) channel
  renormalized two-loop virtual amplitudes for different powers
  of the number of light-quark flavors $\nl$. The normalization is
  defined in Eqs.~(\ref{eq:ampnorm},\ref{eq:ampdecom}) and the scale
  has been set to $\mu = m$.}
\end{figure}

As explained in the previous section, our final results are given for
the finite remainders of the renormalized virtual amplitudes with
$\nl$ active flavors. The results are normalized in such a way that
they directly give the contribution to the respective cross
section. In particular we define
\begin{eqnarray}
\label{eq:ampnorm}
{\cal A}^{(g)}(\beta, \cos\theta) &=& \frac{\beta(1-\beta^2)}{4096\pi}
2 \, \mbox{Re} \, 
  \langle {\cal M}^{(0)}_g | {\cal M}^{(2), \, fin}_g \rangle \; ,
\nonumber \\ \nonumber \\
{\cal A}^{(q)}(\beta,\cos\theta) &=& \frac{\beta(1-\beta^2)}{576\pi} 2
\, \mbox{Re} \, 
  \langle {\cal M}^{(0)}_q | {\cal M}^{(2) , \, fin}_q \rangle \; .
\end{eqnarray}
In both channels, gluon-fusion and quark-annihilation, there is
\begin{equation}
\hat\sigma^{(2)}(\beta) = \frac{\as^4}{m^2} \, \frac{1}{2} \int
d\cos\theta \, {\cal A}(\beta, \cos\theta) \; ,
\end{equation}
where $\hat\sigma^{(2)}(\beta)$ is the two-loop contribution to the
cross section.  Since our results might also be used for other
processes, such a b-quark pair-production, we keep the dependence on
the number of active flavors exact, and further decompose
\begin{equation}
\label{eq:ampdecom}
{\cal A} = {\cal A}_0 + \nl \, {\cal A}_1 + \nl^2 \, {\cal A}_2 \; .
\end{equation}
The values of ${\cal A}^{(g)}$ and ${\cal A}^{(q)}$ on the grid are
attached to this publication in electronic form. We give five
significant digits for each phase space point. This is sufficient for
any practical applications, even if interpolation errors and possible
cancellations with the one-loop-squared contributions are taken into
account. In order to illustrate the complexity of the amplitudes, we
plot them in Fig.~\ref{fig:amplitudes}.

\begin{figure}[h]
\begin{center}
\includegraphics[width=4cm]{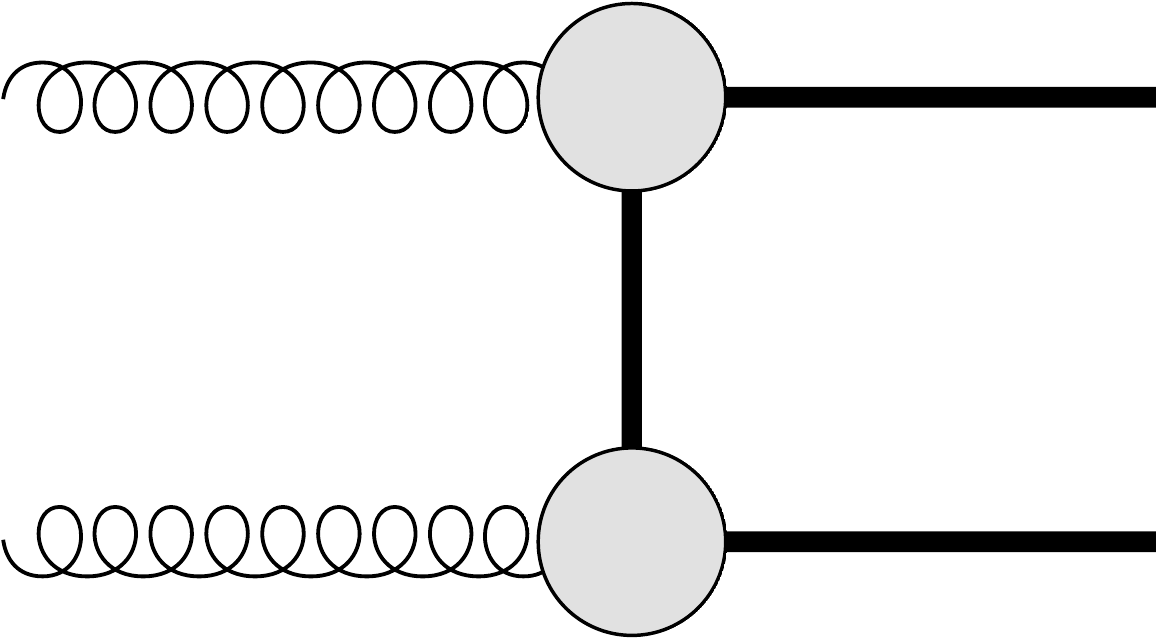}
\end{center}
\caption{\label{fig:tuenhancement} \sf A general diagram responsible for
  the singular behavior of the gluon-fusion amplitude at high energy
  and low scattering angle.}
\end{figure}

As expected, the functions are very smooth. In the case of gluons, we
observe additional enhancements at $\beta \approx 1$ and $\cos\theta
\approx \pm 1$. For low scattering angle, these are due to diagrams of
the general form depicted in Fig.~\ref{fig:tuenhancement}, which
contain a $t$-channel heavy-quark propagator. At tree-level, for
instance, we have
\begin{equation}
\langle {\cal M}_g^{(0)} | {\cal M}_g^{(0)} \rangle \approx 2
\frac{(N_c^2-1)^2}{N_c} \frac{s}{t} \; .
\end{equation}
The behavior at large angles is obtained by $t \leftrightarrow u$
symmetry.

As in previous publications on the subject, we also provide
high-precision values for the color coefficients,
Eqs.~(\ref{eq:colordecgg},\ref{eq:colordecqq}), of the amplitudes at a 
benchmark point. They can be found in Tab.~\ref{tab:ampvalues} for the
  renormalized two-loop virtual amplitudes with $\nl + \nh$ active
  flavors. The numbers for the quark-annihilation channel have been
  presented previously in \cite{Czakon:2008zk}. The table also
  shows, which coefficients are already known analytically. In
  Tab.~\ref{tab:finampvalues}, we present the results at the same
  phase space point, but for the finite remainders with $\nl$ active
  flavors.

\begin{table}
\begin{center}
\begin{tabular}{||l|r|r|r|r|r||}
\hline
\hline
&\multicolumn{1}{c|}{$\ep^{-4}$}&\multicolumn{1}{c|}{$\ep^{-3}$}&\multicolumn{1}{c|}{$\ep^{-2}$}&\multicolumn{1}{c|}{$\ep^{-1}$}&\multicolumn{1}{c||}{$\ep^{0}$}\\
\hline
\hline
$\mathrm{A}^{(g)}$&$10.74942557$&$18.69389337$&$-156.8237244$&$262.1482588$&$12.72180680$\\
$\mathrm{B}^{(g)}$&$-21.28599123$&$-55.99039551$&$-235.0412564$&$1459.833288$&\cellcolor{gray!40}$-509.6019155$\\
$\mathrm{C}^{(g)}$&&$-6.199051597$&$-68.70297402$&$-268.1060373$&\cellcolor{gray!40}$804.0981895$\\
$\mathrm{D}^{(g)}$&&&$94.08660818$&$-130.9619794$&\cellcolor{gray!40}$-283.3496755$\\
$\mathrm{E}^{(g)}_l$&&$-12.54099650$&$18.20646589$&$27.95708293$&$-112.6060988$\\
$\mathrm{E}^{(g)}_h$&&&$0.012907497$&$11.79259573$&\cellcolor{gray!40}$-47.68412574$\\
$\mathrm{F}^{(g)}_l$&&$24.83365643$&$-26.60868620$&$-50.75380859$&$125.0537955$\\
$\mathrm{F}^{(g)}_h$&&&$0.0$&$-23.32918072$&\cellcolor{gray!40}$132.5618962$\\
$\mathrm{G}^{(g)}_l$&&&$3.099525798$&$67.04300456$&$-214.1081462$\\
$\mathrm{G}^{(g)}_h$&&&&$0.0$&\cellcolor{gray!40}$-179.3374874$\\
$\mathrm{H}^{(g)}_l$&&&$2.388761238$&$-5.452031425$&$3.632861953$\\
$\mathrm{H}^{(g)}_{lh}$&&&&$-0.004302499$&$-3.945712447$\\
$\mathrm{H}^{(g)}_h$&&&&&\cellcolor{gray!40}$0.00439856$\\
$\mathrm{I}^{(g)}_l$&&&$-4.730220272$&$10.81032548$&$-7.182940516$\\
$\mathrm{I}^{(g)}_{lh}$&&&&$0.0$&$7.780900470$\\
$\mathrm{I}^{(g)}_h$&&&&&\cellcolor{gray!40}$0.0$\\
\hline
\hline
$\mathrm{A}^{(q)}$&$0.22625$&$1.391733154$&$-2.298174307$&$-4.145752449$&$17.37136599$\\
$\mathrm{B}^{(q)}$&$-0.4525$&$-1.323646320$&$8.507455541$&$6.035611156$&\cellcolor{gray!40}$-35.12861106$\\
$\mathrm{C}^{(q)}$&$0.22625$&$-0.06808683395$&$-18.00716652$&$6.302454931$&\cellcolor{gray!40}$3.524044913$\\
$\mathrm{D}^{(q)}_l$&&$-0.22625$&$0.2605057339$&$-0.7250180282$&$-1.935417247$\\
$\mathrm{D}^{(q)}_h$&&&$0.5623350684$&$0.1045606449$&$-1.704747998$\\
$\mathrm{E}^{(q)}_l$&&$0.22625$&$-0.3323207300$&$7.904121951$&$2.848697837$\\
$\mathrm{E}^{(q)}_h$&&&$-0.5623350684$&$4.528240788$&$12.73232424$\\
$\mathrm{F}^{(q)}_l$&&&&&$-1.984228442$\\
$\mathrm{F}^{(q)}_{lh}$&&&&&$-2.442562819$\\
$\mathrm{F}^{(q)}_h$&&&&&$-0.07924540546$\\
\hline
\hline
\end{tabular}
\caption{\label{tab:ampvalues} \sf Values of the color coefficients
  defined in Eqs.~(\ref{eq:colordecgg},\ref{eq:colordecqq}) of the
  renormalized two-loop virtual amplitudes with $\nl + \nh$ active
  flavors for the gluon-fusion and quark-annihilation channels at
  $m^2/s = 0.2$, $t/s = 0.45$ with $\mu = m$.  The entries with
  explicit $0.0$ vanish because of the scale choice. The gray shaded
  values are only known numerically, whereas the remaining ones have
  been obtained analytically in \cite{Ferroglia:2009ii,
    Bonciani:2008az, Bonciani:2009nb, Bonciani:2010mn,
    vonManteuffel:2013uoa, Bonciani:2013ywa}.}
\end{center}
\end{table}

\begin{table}
\begin{center}
\begin{tabular}{||l|r|r||}
\hline
\hline
&\multicolumn{1}{c|}{$gg \to Q\bar Q$}&\multicolumn{1}{c||}{$q
  \overline{q} \to Q\bar Q$}\\
\hline
\hline
A&$99.35775524$&$18.51259223$\\
B&$50.28087862$&$-30.28872915$\\
C&$1139.719830$&$-24.73812607$\\
D&$24.99913023$&\\
\hline
$\mathrm{D}_l$&&$-1.473159900$\\
$\mathrm{E}_l$&$-39.12377988$&$10.41179740$\\
$\mathrm{F}_l$&$64.56254598$&$-1.984228442$\\
$\mathrm{G}_l$&$-172.6559290$&\\
$\mathrm{H}_l$&$-0.0004110707375$&\\
$\mathrm{I}_l$&$0.0$&\\
\hline
$\mathrm{D}_h$&&$0.7062093236$\\
$\mathrm{E}_h$&$-11.88421959$&$6.003261199$\\
$\mathrm{F}_h$&$51.95989356$&$-0.07924540546$\\
$\mathrm{G}_h$&$-136.0272435$&\\
$\mathrm{H}_h$&$0.004398562176$&\\
$\mathrm{I}_h$&$0.0$&\\
\hline
$\mathrm{F}_{lh}$&&$-2.442562819$\\
$\mathrm{H}_{lh}$&$-0.01892800046$&\\
$\mathrm{I}_{lh}$&$0.0$&\\
\hline
\hline
\end{tabular}
\caption{\label{tab:finampvalues} \sf Values of the color coefficients
  defined in Eqs.~(\ref{eq:colordecgg},\ref{eq:colordecqq}) of the
  finite remainders of the renormalized two-loop virtual amplitudes
  with $\nl$ active flavors for the gluon-fusion (second column) and
  quark-annihilation (third column) channels at $m^2/s = 0.2$, $t/s = 0.45$
  with $\mu = m$. The entries with explicit $0.0$ vanish because of
  the scale choice.}
\end{center}
\end{table}

While the grid covers a substantial part of the phase space, it
cannot completely cover the high-energy and threshold domains, since
the amplitudes are singular in these limits. For these two cases, we
provide expansions. The leading behavior of the high-energy
expansions for the renormalized amplitudes has been first determined
in Ref.~\cite{Czakon:2007ej, Czakon:2007wk}. We have calculated the
first three terms of the expansions of the bare two-loop amplitudes
analytically and converted them into finite remainders. The results
are attached to this publication in electronic form. Notice, that the
results for each term of the expansions are exact in the scattering
angle. However, they are in principle only valid if $t,u \gg
m^2$, since the expansion is really in $m^2/s$, $m^2/t$, and $m^2/u$,
under the assumption $t/s, \, u/s, \, t/u = {\cal O}(1)$.

The question of the validity of the high-energy expansions at
small/large scattering angle deserves a more careful study. We discuss
the small scattering angle case only, since the opposite limit is
analogous. First, we note that at $\cos\theta = 1$, there is $t/s =
1/2(1 - \beta) > m^2/s$, with $t/s \to m^2/s$ for $m^2/s \to 0$. From
this, we conclude that the ratio $m^2/t$ should be considered of the
order of unity in the high-energy low-scattering-angle limit. The
convergence of the expansions depends, therefore, on the coefficients of
the series. By inspection, we note that the expansions seem to have a
non-zero radius of convergence at $\cos\theta = 1$.

Independently of these considerations, in the high-energy small/large
scattering-angle case, cross sections are dominated by real radiation,
which means that the issue of the behavior of the virtual
amplitudes in this region is moot.

In the threshold region, the amplitudes are dominated by singularities
coming from potential interactions between the heavy quarks. Beyond
the grid, sufficient precision is obtained already with the leading
behavior up to terms of ${\cal O}(\beta^0)$. For the application of
our results to threshold expansions of cross sections, we are also
interested in decomposing the gluon-fusion channel amplitudes into
color structures. In general, there are three SU(3) invariant color
structures linking two incoming gluons with color indices $a$ and $b$,
with a final state heavy-quark with color index $c$ and anti-quark
with color index $d$: 1) color singlet, {\bf 1}, given by
$\delta^{ab}\delta_{cd}$; 2) symmetric color octet, ${\bf 8_S}$, given
by $d^{abe}T^e_{cd}$, with $d^{abc} = 2 \, {\rm Tr}
(T^a T^b T^c + T^c T^b T^a)$; and 3) anti-symmetric color octet, ${\bf
  8_A}$, given by $i f^{abe}T^e_{cd}$. For each of these structures,
denoted by $\alpha$, we introduce a color projector, ${\cal
  P}_\alpha$, such that ${\cal P}^2_\alpha = {\cal P}_\alpha$, and
${\cal P}_{\bf 1} + {\cal P}_{\bf 8_S} + {\cal P}_{\bf 8_A} = 1$. The
expansions of the finite remainders of the renormalized two-loop
amplitudes with $\nl$ active flavors at $\mu = m$ read
\begin{eqnarray}
\label{eq:expg1}
&& 2 \, \mbox{Re} \, \langle {\cal M}_g^{(0)} | {\cal P}_{\bf 1}| {\cal
  M}_g^{(2), \, fin} \rangle =
\langle {\cal M}_g^{(0)} | {\cal P}_{\bf 1}| {\cal
  M}_g^{(0)} \rangle \, \cf \, \pi^2 \,
\nonumber \\ && \hspace{1cm} \times
\biggl\{
- \frac{1}{\beta^2} \cf \biggl[ \lb2 + 2 \lt1 \lb1 + \lt2
-\frac{\pi^2}{12} \biggr]
+ \frac{1}{\beta} \biggl[
\biggl(
\ca \biggl( -\frac{11}{3} -4 \lt1 \biggr)
+ \frac{2}{3} \nl \biggr) \lb1 
\nonumber \\ && \hspace{1.6cm}
+ \ca \biggl( \frac{49}{18} - \frac{11}{3} \lt1 -6 \lt2
+\frac{\pi^2}{3} \biggr)
+ \cf \biggl( -5 + \frac{\pi^2}{4} \biggr)
+ \nl \biggl( - \frac{5}{9} + \frac{2}{3} \lt1 \biggr)
\biggr]
\nonumber \\ && \hspace{1.6cm}
- 4 \cf \lb2 - \biggl( 4 \ca + \cf (4 + 8 \lt1) + \cf \cos^2\theta
\biggr) \lb1
\biggr\} 
\nonumber \\ && \hspace{1cm}
+ \Delta^{(g, \bf 1)}_{0} + \Delta^{(g, \bf 1)}_{2}
\cos^2\theta \; ,
\\ \nonumber \\
\label{eq:expg8S}
&& 2 \, \mbox{Re} \, \langle {\cal M}_g^{(0)} | {\cal P}_{\bf 8_S}| {\cal
  M}_g^{(2), \, fin} \rangle =
\langle {\cal M}_g^{(0)} | {\cal P}_{\bf 8_S}| {\cal
  M}_g^{(0)} \rangle \, \left( \cf -\frac{\ca}{2} \right) \, \pi^2 \,
\nonumber \\ && \hspace{1cm} \times
\biggl\{
- \frac{1}{\beta^2} \biggl( \cf - \frac{\ca}{2} \biggr) \biggl[ \lb2 +
  2 \lt1 \lb1 + \lt2 -\frac{\pi^2}{12} \biggr]
+ \frac{1}{\beta} \biggl[
\biggl( \ca \biggl( -\frac{11}{3} -2 \lt1 \biggr)
\nonumber \\ && \hspace{1.6cm}
+ \frac{2}{3} \nl
\biggr) \lb1
+ \ca \biggl( \frac{67}{18} - \frac{8}{3} \lt1 -4 \lt2
+\frac{5\pi^2}{24} \biggr)
+\cf \biggl( -5 + \frac{\pi^2}{4} \biggr)
\nonumber \\ && \hspace{1.6cm}
+ \nl \biggl( -\frac{5}{9} + \frac{2}{3} \lt1 \biggr)
\biggr]
- 4 \biggl( \cf - \frac{\ca}{2} \biggr) \lb2
- \biggl( \ca (2 - 4 \lt1 ) + \cf ( 4 + 8 \lt1) 
\nonumber \\ && \hspace{1.6cm}
+ \biggl( \cf - \frac{\ca}{2} \biggr) \cos^2\theta \biggr) \lb1
\biggr\}
\nonumber \\ && \hspace{1cm}
+ \Delta^{(g, \bf 8_S)}_{0} + \Delta^{(g, \bf 8_S)}_{2}
\cos^2\theta \; ,
\\ \nonumber \\
&& 2 \, \mbox{Re} \, \langle {\cal M}_g^{(0)} | {\cal P}_{\bf 8_A}| {\cal
  M}_g^{(2), \, fin} \rangle = 
\langle {\cal M}_g^{(0)} | {\cal P}_{\bf 8_A}| {\cal
  M}_g^{(0)} \rangle \, \left( \cf -\frac{\ca}{2} \right)^2 \, \pi^2
\, \nonumber \\ && \hspace{1cm} \times
\frac{1}{\beta^2} \biggl[ -\lb2 + 2 (1 - \lt1) \lb1 + \lt1 (2 - \lt1) +
\frac{\pi^2}{12} \biggr] \; ,
\\ \nonumber \\
\label{eq:expq}
&& 2 \, \mbox{Re} \, \langle {\cal M}_q^{(0)} | {\cal
  M}_q^{(2), \, fin} \rangle =
\langle {\cal M}_q^{(0)} | {\cal M}_q^{(0)} \rangle \, \left( \cf
-\frac{\ca}{2} \right) \, \pi^2 \, \nonumber \\ && \hspace{1cm} \times
\biggl\{
- \frac{1}{\beta^2} \biggl( \cf - \frac{\ca}{2} \biggr) \biggl[ \lb2 +
  2 \lt1 \lb1 + \lt2 -\frac{\pi^2}{12} \biggr]
+ \frac{1}{\beta} \biggl[
\biggl( \ca \biggl( -\frac{31}{6} + 2 \lt1 \biggr)
\nonumber \\ && \hspace{1.6cm}
+\cf (3 - 4 \lt1) +\frac{4}{3} \nl
\biggr) \lb1
+ \ca \biggl( \frac{131}{18} - \frac{43}{6} \lt1 + 2 \lt2 -
\frac{\pi^2}{4} \biggr)
\nonumber \\ && \hspace{1.6cm}
+ \cf \biggl( -8 + 6 \lt1 - 6 \lt2 +\frac{7\pi^2}{12} \biggr)
- \frac{8}{9} \nh + \nl \biggl( -\frac{10}{9} + 2 \lt1 \biggr)
\biggr]
\nonumber \\ && \hspace{1.6cm}
- \frac{1}{2} \biggl( \cf - \frac{\ca}{2} \biggr) ( 3 + \cos^2\theta )
\lb2 
+ \biggl( \ca \biggl( -\frac{41}{12} + \frac{3}{2} \lt1 \biggr) 
+ \cf \biggl( -\frac{7}{6} - 3 \lt1 \biggr) 
\nonumber \\ && \hspace{1.6cm}
+ 3 \nh
+ \biggl( -2 \ca + \frac{16}{3} \cf \biggr) \cos\theta
+ \frac{1}{2} \biggl( \cf - \frac{\ca}{2} \biggr) ( 1 - 2 \lt1 )
\cos^2\theta \biggr) \lb1
\biggr\}
\nonumber \\ && \hspace{1cm}
+ \Delta^{(q)}_{0} + \Delta^{(q)}_{1} \cos\theta + \Delta^{(q)}_{2}
\cos^2\theta \; ,
\end{eqnarray}
where the Born matrix elements near threshold are given by
\begin{eqnarray}
\label{eq:BornMatrix}
&& \langle {\cal M}_g^{(0)} | {\cal P}_{\bf 1}| {\cal
  M}_g^{(0)} \rangle = 4 \, \frac{N_c^2-1}{N_c} \; ,
\nonumber \\ \nonumber \\
&& \langle {\cal M}_g^{(0)} | {\cal P}_{\bf 8_S}| {\cal
  M}_g^{(0)} \rangle = 2 \frac{(N_c^2-1)(N_c^2-4)}{N_c} \; ,
\nonumber \\ \nonumber \\
&& \langle {\cal M}_g^{(0)} | {\cal P}_{\bf 8_A}| {\cal
  M}_g^{(0)} \rangle = 2 N_c(N_c^2 - 1) \beta^2 \cos^2\theta \; ,
\nonumber \\ \nonumber \\
&& \langle {\cal M}_q^{(0)} | {\cal M}_q^{(0)} \rangle = 2(N_c^2 - 1)
\; .
\end{eqnarray}

\begin{table}
\begin{center}
\begin{tabular}{||l|r|r|r|r||}
\hline
\hline
&\multicolumn{1}{c|}{$\Delta^{(g, \bf
    1)}_{0}$}&\multicolumn{1}{c|}{$\Delta^{(g, \bf
    1)}_{2}$}&\multicolumn{1}{c|}{$\Delta^{(g, \bf
    8_S)}_{0}$}&\multicolumn{1}{c|}{$\Delta^{(g, \bf 8_S)}_{2}$}\\
\hline
\hline
$\mathrm{A}^{(g)}$&&&$12.64743334$&\\
$\mathrm{B}^{(g)}$&$-136.8630600$&$-16.71069286$&$-8.317638322$&\\
$\mathrm{C}^{(g)}$&$25.21974837$&$33.42138573$&$-179.8492935$&$-8.355346432$\\
$\mathrm{D}^{(g)}$&$-21.52182675$&$-16.71069286$&$43.04365351$&$33.42138573$\\
$\mathrm{E}^{(g)}_l$&&&$-5.762337505$&\\
$\mathrm{E}^{(g)}_h$&&&$1.250977414$&\\
$\mathrm{F}^{(g)}_l$&$0.1795129111$&&$19.45833374$&\\
$\mathrm{F}^{(g)}_h$&$6.632216653$&&$-11.07432229$&\\
$\mathrm{G}^{(g)}_l$&$-1.331755821$&&$14.36406511$&\\
$\mathrm{G}^{(g)}_h$&$-6.965100683$&&$24.28165052$&\\
\hline
\hline
\end{tabular}
\caption{\label{tab:constgg} \sf Values of the constant coefficients
  of the $\beta$-expansions, Eqs.~(\ref{eq:expg1}, \ref{eq:expg8S}), of
  the finite remainder of the renormalized two-loop virtual amplitude
  with $\nl$ active flavors in the gluon-fusion channel for the color
  singlet, $\Delta^{(g,\bf 1)}$, and symmetric octet, $\Delta^{(g,\bf
    8_S)}$, at $\mu = m$. The coefficients of the unlisted entries
  vanish.}
\end{center}
\end{table}

\begin{table}
\begin{center}
\begin{tabular}{||l|r|r|r||}
\hline
\hline
&\multicolumn{1}{c|}{$\Delta^{(q)}_0$}&\multicolumn{1}{c|}{$\Delta^{(q)}_1$}&\multicolumn{1}{c|}{$\Delta^{(q)}_2$}\\
\hline
\hline
$\mathrm{A}^{(q)}$&$21.39026702$&&\\
$\mathrm{B}^{(q)}$&$41.88839539$&$-19.23445615$&\\
$\mathrm{C}^{(q)}$&$-53.87050002$&$76.93782458$&$0.8658454253$\\
$\mathrm{D}^{(q)}_l$&$-1.736456791$&&\\
$\mathrm{D}^{(q)}_h$&$-2.451777776$&&\\
$\mathrm{E}^{(q)}_l$&$-0.5233480317$&&\\
$\mathrm{E}^{(q)}_h$&$-13.46305281$&&\\
$\mathrm{F}^{(q)}_l$&$-2.175776838$&&\\
$\mathrm{F}^{(q)}_{lh}$&$0.3322931029 $&&\\
$\mathrm{F}^{(q)}_h$&$1.580246914$&&\\
\hline
\hline
\end{tabular}
\caption{\label{tab:constqq} \sf Values of the constant coefficients
  of the $\beta$-expansion, Eq.~(\ref{eq:expq}), of the finite
  remainder of the renormalized two-loop virtual amplitude with $\nl$
  active flavors in the quark-annihilation channel with $\mu = m$.} 
\end{center}
\end{table}

The formulae for the expansions of the dominant channels contain
constants, which we have only determined numerically. They are given in
Tabs.~\ref{tab:constgg},~\ref{tab:constqq}. Notice that several
coefficients in the second table (of fermionic and leading-color
bosonic origin) could be determined analytically using the results of
Ref.~\cite{Bonciani:2008az, Bonciani:2009nb}. This requires, however,
to add analytic results for the infrared counter-terms that transform
renormalized amplitudes into finite remainders. In the gluon channel,
analytic results are only known for the sum of the color-structure
projected amplitudes. It is thus not possible to obtain any
coefficients presented in Tab.~\ref{tab:constgg} in analytic form with
the currently available results from the literature.


\section{Threshold expansions of cross sections}
\label{sec:ThCross}

Our results for the two-loop virtual amplitudes can be used to obtain
the leading threshold behavior of partonic top-quark pair-production
cross sections at next-to-next-to-leading order. When $\beta \approx
0$, cross sections are dominated by terms of the form $\beta \times
1/\beta^n \ln^m\beta$ with $n > 0$ and/or $m > 0$. While the first factor
of $\beta$ is a phase space suppression, the enhancements by positive
powers of $\ln\beta$ and $1/\beta$ are due  to emissions of soft
gluons and non-relativistic potential interactions between the quark
and the anti-quark. At next-to-next-to-leading order, the coefficients
of these singular terms have been determined in
Ref.~\cite{Beneke:2009ye}. We would like to extend this analysis to
include terms with $n = m = 0$, which are velocity independent
with respect to the Born cross section. The threshold expansion
including both singular and constant terms can be obtained from
factorization as explained in Ref.~\cite{Czakon:2013hxa}. According to
the latter publication, close to threshold, a cross section for a
given initial state can be written as
\begin{equation}
\label{eq:factorization}
\hat\sigma = \sum_\alpha H_\alpha \otimes S_\alpha \; .
\end{equation}
$H_\alpha$ are called hard functions, and are obtained by
expanding in $\beta$ the partonic cross sections obtained exclusively
with the finite remainder of the virtual amplitudes projected onto the
color configuration $\alpha$. Therefore, $H_\alpha$ do not contain any
real-radiation effects. $S_\alpha$ are called soft functions, and are
given by cross sections for emission of gluons and light-quark pairs
from eikonal lines representing the external partons of the hard
process in the color configuration $\alpha$. The convolution is
performed in the energy of the soft radiation. The color
configurations, $\alpha$, must correspond to irreducible
representations of the SU(3) group, in order for this simple form to
be valid. The cross section expansions are thus labeled by the color
configuration. Accordingly, we introduce the perturbative expansions
\begin{eqnarray}
\hat\sigma_{ij,\alpha}(\beta,\mu,m) &=& \hat\sigma^{(0)}_{ij,\alpha} \Bigg\{ 1
+ \frac{\as(\mu^2)}{4\pi} \left[\hat\sigma^{(1,0)}_{ij,\alpha} +
\hat\sigma^{(1,1)}_{ij,\alpha}\ln\left({\mu^2\over m^2}\right) \right]
\label{eq:result-scales}\\
&&+\left(\frac{\as(\mu^2)}{4\pi}\right)^2
\left[\hat\sigma^{(2,0)}_{ij,\alpha} +
  \hat\sigma^{(2,1)}_{ij,\alpha}\ln\left({\mu^2\over m^2}\right) +
  \hat\sigma^{(2,2)}_{ij,\alpha}\ln^2\left({\mu^2\over m^2}\right)\right]
+ {\cal O}(\as^3) \Bigg\} \; ,\nonumber
\end{eqnarray}
where we only consider the initial states $ij = q\bar{q}, gg$, and the
possible color configurations are $\alpha \in \{ {\bf 1}, {\bf 8} \}$
for the quark-annihilation channel, while $\alpha \in \{ {\bf 1}, {\bf
  8_S}, {\bf 8_A} \}$ for the gluon-fusion channel. In this section, we work
exclusively with $\nl$ active flavors, which implies that $\as =
\asnl$. We will again neglect the scale dependence and set $\mu = m$
in the subsequent expressions.

Using the matrix elements Eq.~(\ref{eq:BornMatrix}), we obtain the
following threshold expansions for the dominant color-projected Born
cross sections in the gluon-fusion channel
\begin{equation}
\hat\sigma^{(0)}_{gg,{\bf 1}} = \pi\beta{1\over 4N_c(N_c^2-1)}
{\as^2\over m^2} \, , \;\;\;\;
\hat\sigma^{(0)}_{gg,{\bf 8_S}} = \pi\beta{(N_c^2-4)\over 8N_c(N_c^2-1)}
{\as^2\over m^2} \, .
\end{equation}
As was pointed out in \cite{Beenakker:2013mva}, it is a coincidence
that the anti-symmetric octet cross section is sub-leading in the
threshold region. This fact cannot be justified with symmetry
arguments (the often quoted Landau-Yang theorem does not apply to
gluons). There is
\begin{equation}
\hat\sigma^{(0)}_{gg,{\bf 8_A}} = \pi\beta^3{N_c\over 24(N_c^2-1)}
  {\as^2\over m^2} \, .
\end{equation}
Finally, the quark channel receives octet contributions only
\begin{equation}
\hat \sigma_{q\bar{q}}^{(0)} = \hat \sigma_{q\bar{q}, {\bf 8}}^{(0)}
= \pi\beta{(N_c^2-1)\over 8N_c^2}
{\as^2\over m^2} \; .
\end{equation}
Notice that while singlet contributions do not vanish beyond
leading order in this channel, they are not enhanced at threshold and
are of ${\cal O}(\beta^3)$.

We are interested in the threshold expansions of the
next-to-next-to-leading order contributions,
$\hat\sigma^{(2)}_{ij,\alpha}$, which take the form
\begin{equation}
\hat\sigma^{(2)}_{ij,\alpha} = \sum_{n=0}^2 \sum_{m=0}^{4-2n}
c^{(n,m)}_{ij,\alpha}\frac{1}{\beta^n} \ln^m\beta \; .
\end{equation}
The coefficients of the singular terms of the $\beta$-expansions,
$c^{(n,m)}_{ij,\alpha}$ with $n > 0$ and/or $m > 0$, are in principle
known \cite{Beneke:2009ye}, although a contribution has been omitted
in the analysis of $c^{(0,1)}_{q\bar q}$, as we explain in the
following. For future reference, we define
\begin{equation}
C^{(2)}_{ij,\alpha} = c^{(0,0)}_{ij,\alpha} \; ,
\end{equation}
for all but the $ij = gg, \alpha = {\bf 8_A}$ case. For the latter, we
further define
\begin{equation}
C^{(2)}_{gg, \bf 8_A} = \frac{\hat\sigma^{(0)}_{gg, \bf
    8_A}}{\hat\sigma^{(0)}_{gg,\bf 1} + \hat\sigma^{(0)}_{gg,\bf 8_S}}
\, \hat\sigma^{(2)}_{gg,\bf 8_A} \; ,
\end{equation}
where the right hand side has been expanded to ${\cal O}(\beta^0)$.
The coefficient $C^{(2)}_{gg, \bf 8_A}$ is proportional to
$c^{(2,0)}_{gg, \bf 8_A}$.

\begin{figure}
{\center
\begin{tabular}{lll}
\includegraphics[height=3cm]{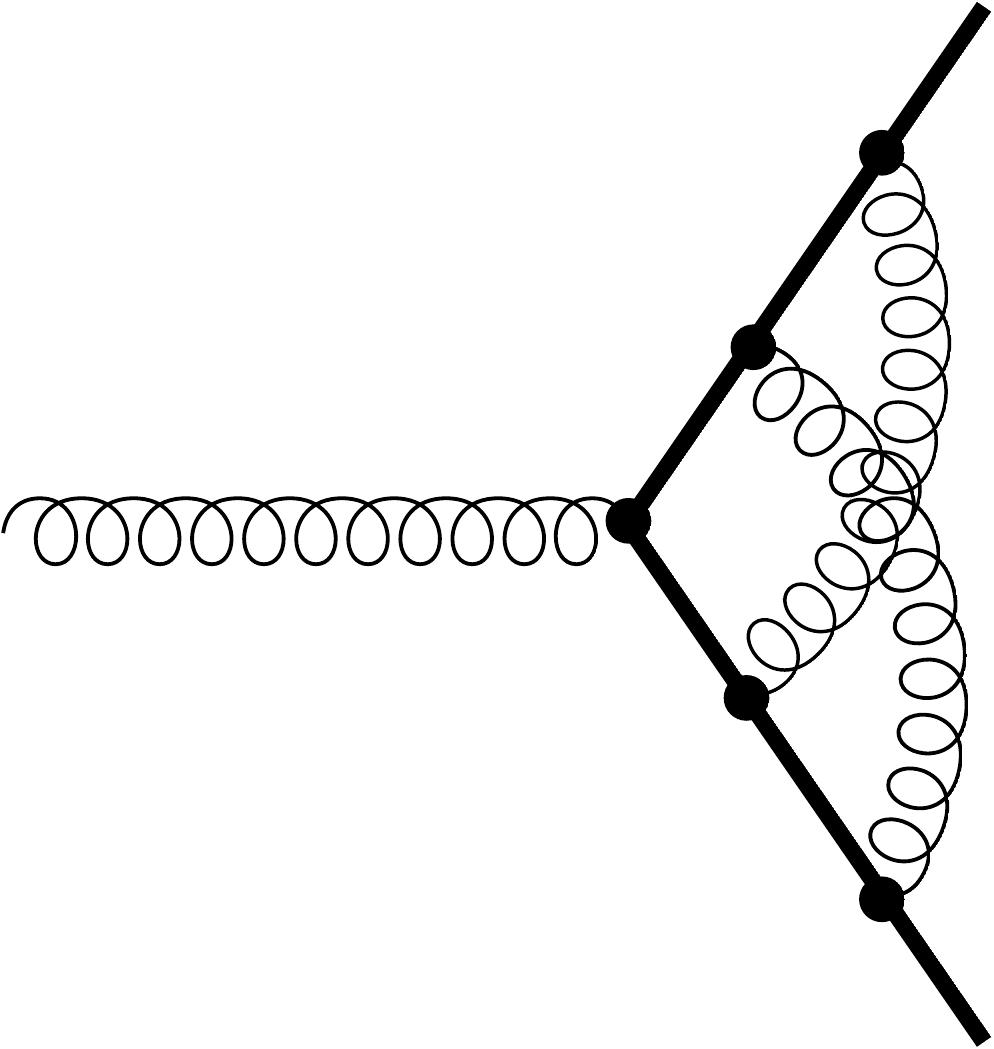} & \hspace{2cm} &
\includegraphics[height=3cm]{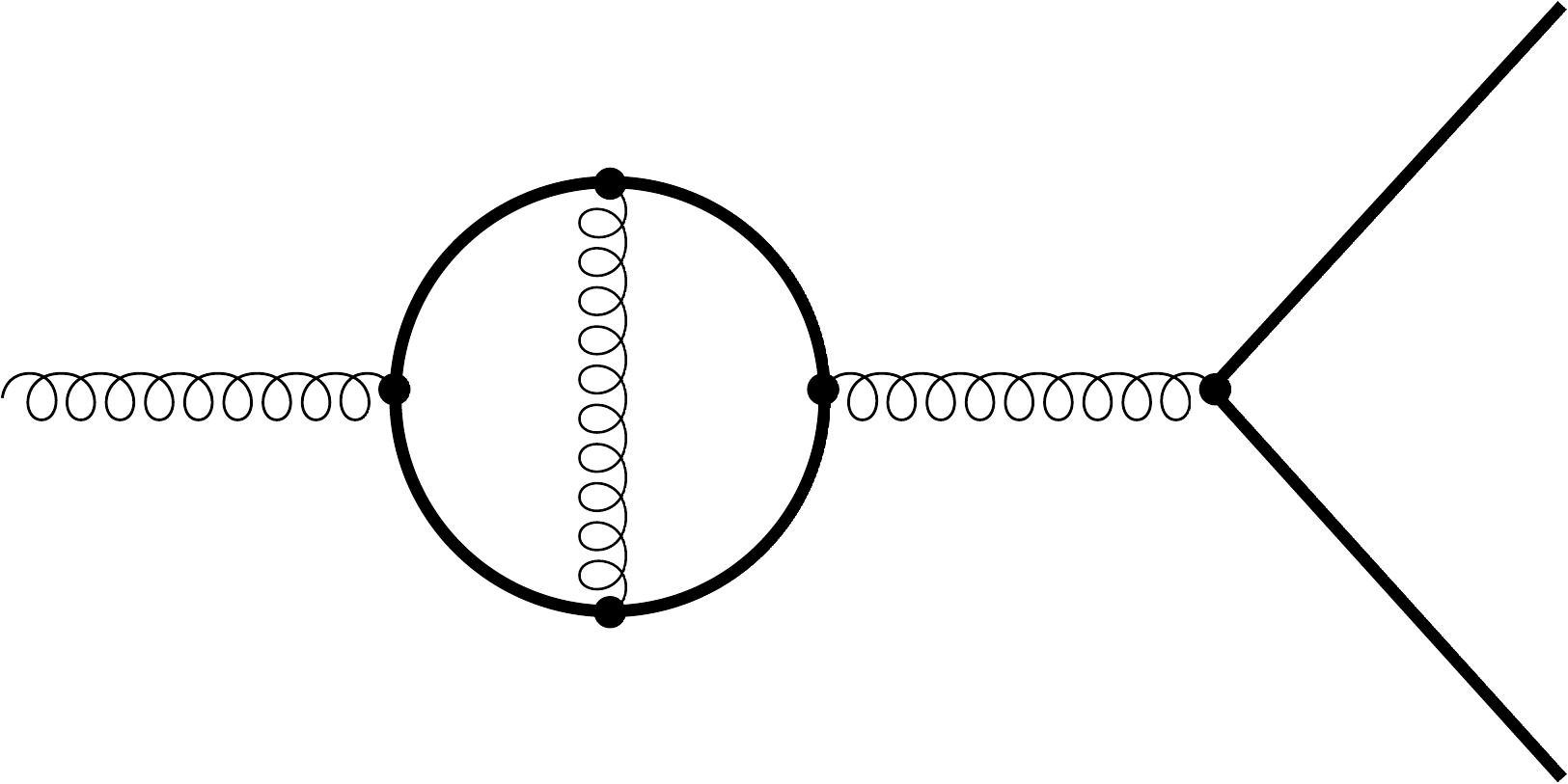} \\
a) & & b)
\end{tabular}
\caption{\label{fig:potential} \sf Examples of two-loop contributions
  leading to a singular dependence of the cross section on the
  heavy-quark (thick line) velocity near threshold.}}
\end{figure}

The purpose of this section is to obtain
$C^{(2)}_{ij,\alpha}$. However, we will first determine a missing
contribution to $c^{(0,1)}_{q\bar q}$. To this end, we return to  the
derivation of the coefficients of the $\beta$-singular terms in
Ref.~\cite{Beneke:2009ye}, which proceeded with an additional
factorization of potential effects. These can, in fact, be
classified into two categories. One category corresponds to the case,
when a quark-anti-quark system is created, and subsequently strongly
interacts through effective potentials, dominated by the Coulomb
potential. An example diagram can be found in Fig.~\ref{fig:potential}
a). This category gives a complete description of the singlet case,
and was the only one taken into account in
Ref.~\cite{Beneke:2009ye}. However, in the case of an octet final
state configuration there is another category of enhanced
contributions. Indeed, the pair can then also annihilate, as shown in
Fig.~\ref{fig:potential} b). Here, the Coulomb exchange between the
virtual heavy quarks leads to a  logarithmic singularity as we will
demonstrate below. Unfortunately, since this type of enhancements has
not been taken into account in Ref.~\cite{Beneke:2009ye}, the cross
section expansions given there do not contain all the singular
terms. The additional logarithm contributes to the quark-annihilation
channel only. The reason for the absence of enhancement in the
gluon-fusion channel is that the diagram of Fig.~\ref{fig:potential}
b) can only occur in the anti-symmetric octet configuration, where the
$s$-channel gluon is produced from a triple-gluon vertex. At
next-to-next-to-leading order, this diagram is contracted with the
Born amplitude, which is sub-leading in the anti-symmetric octet, and
the contribution to the total cross section is only of ${\cal
  O}(\beta^3 \lb1)$.

In principle, we could determine the correction to the singular
expansion terms directly from Eq.~(\ref{eq:factorization}) after
substituting the expansions of the virtual amplitudes from the previous
section. However, the missing term due to Fig.~\ref{fig:potential} b)
can be obtained with simple arguments based on unitarity and
analyticity, which  helps to clarify the origin of the logarithmic
enhancement. We start with the definition of the vacuum polarization
for gluons
\begin{equation}
i \, \Pi^{ab}_{\mu\nu}(q) =
\vcenter{\hbox{\includegraphics[height=1.5cm]{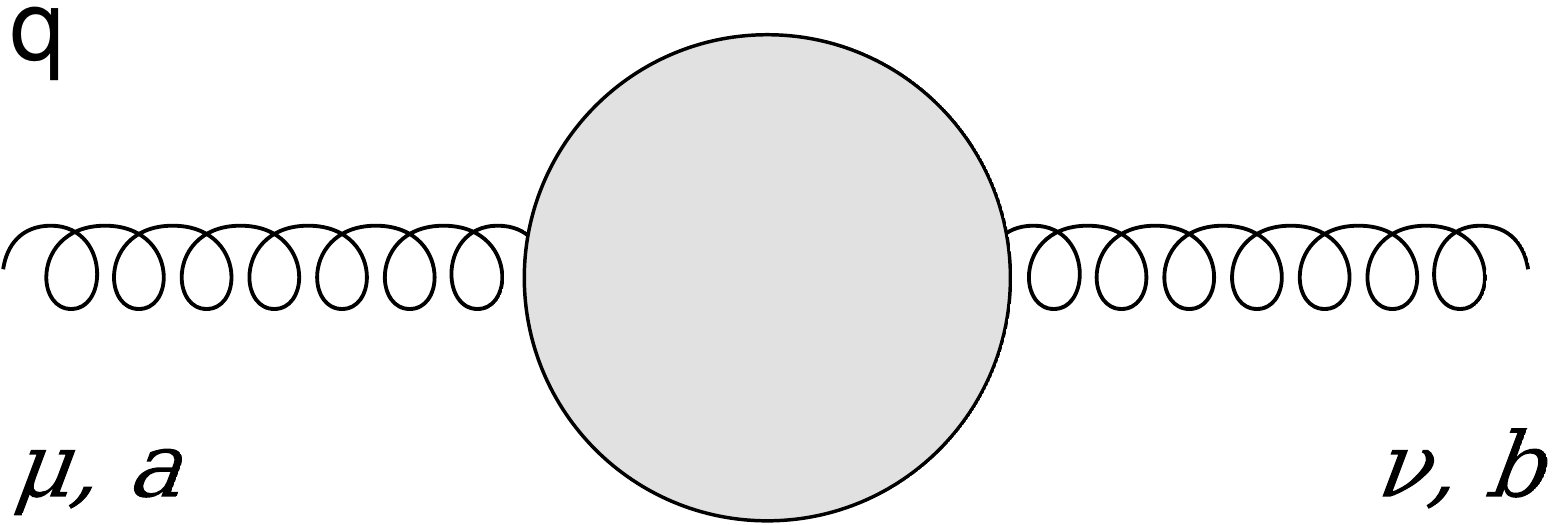}}}
\; ,
\end{equation}
and its decomposition by transversality
\begin{equation}
\Pi^{ab}_{\mu\nu}(q) = (g_{\mu\nu} q^2 - q_\mu q_\nu) \, \delta^{ab} \,
\Pi(q^2) \; .
\end{equation}
With these definitions, it is easy to see that
\begin{equation}
\vcenter{\hbox{\includegraphics[height=2.5cm]{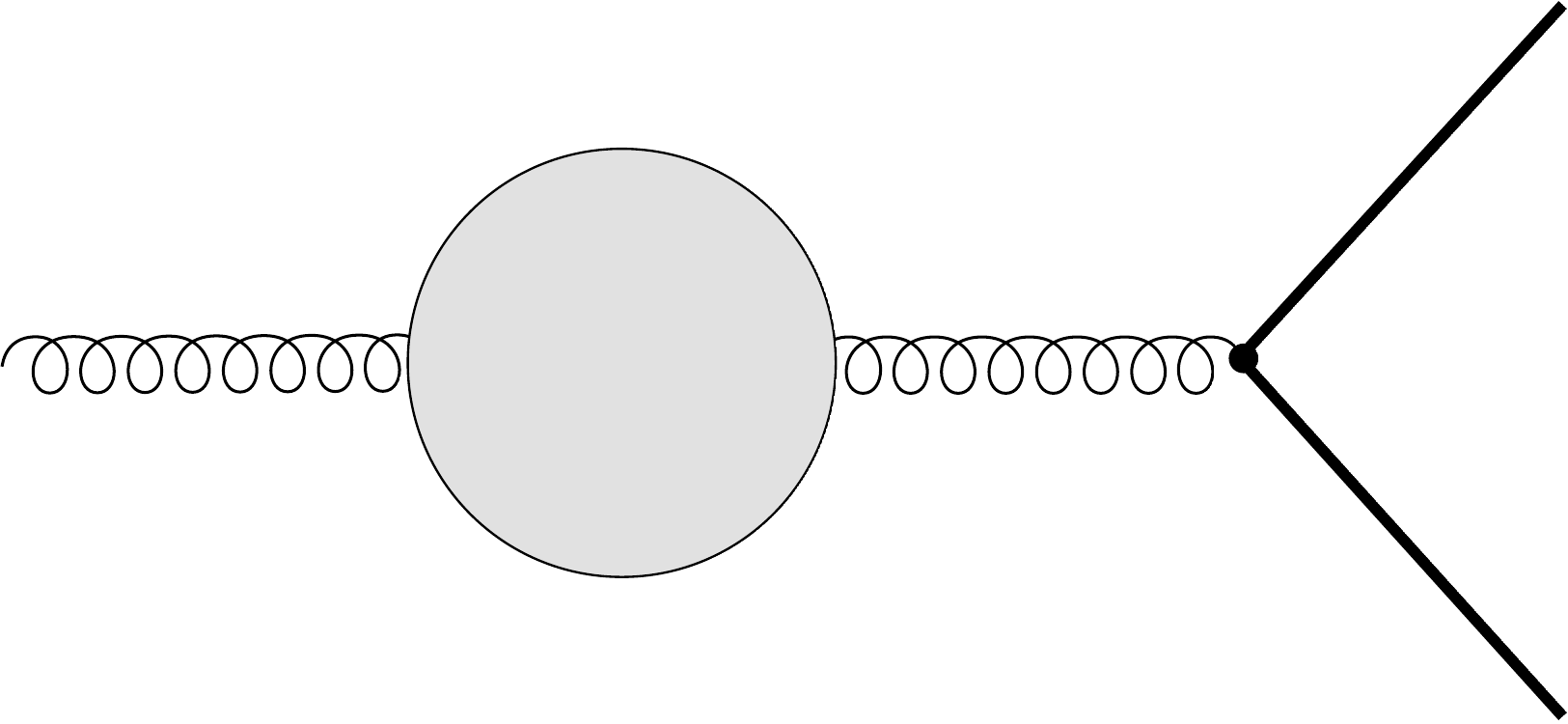}}}
\; = \; \Pi(s) \times
\vcenter{\hbox{\includegraphics[height=2.5cm]{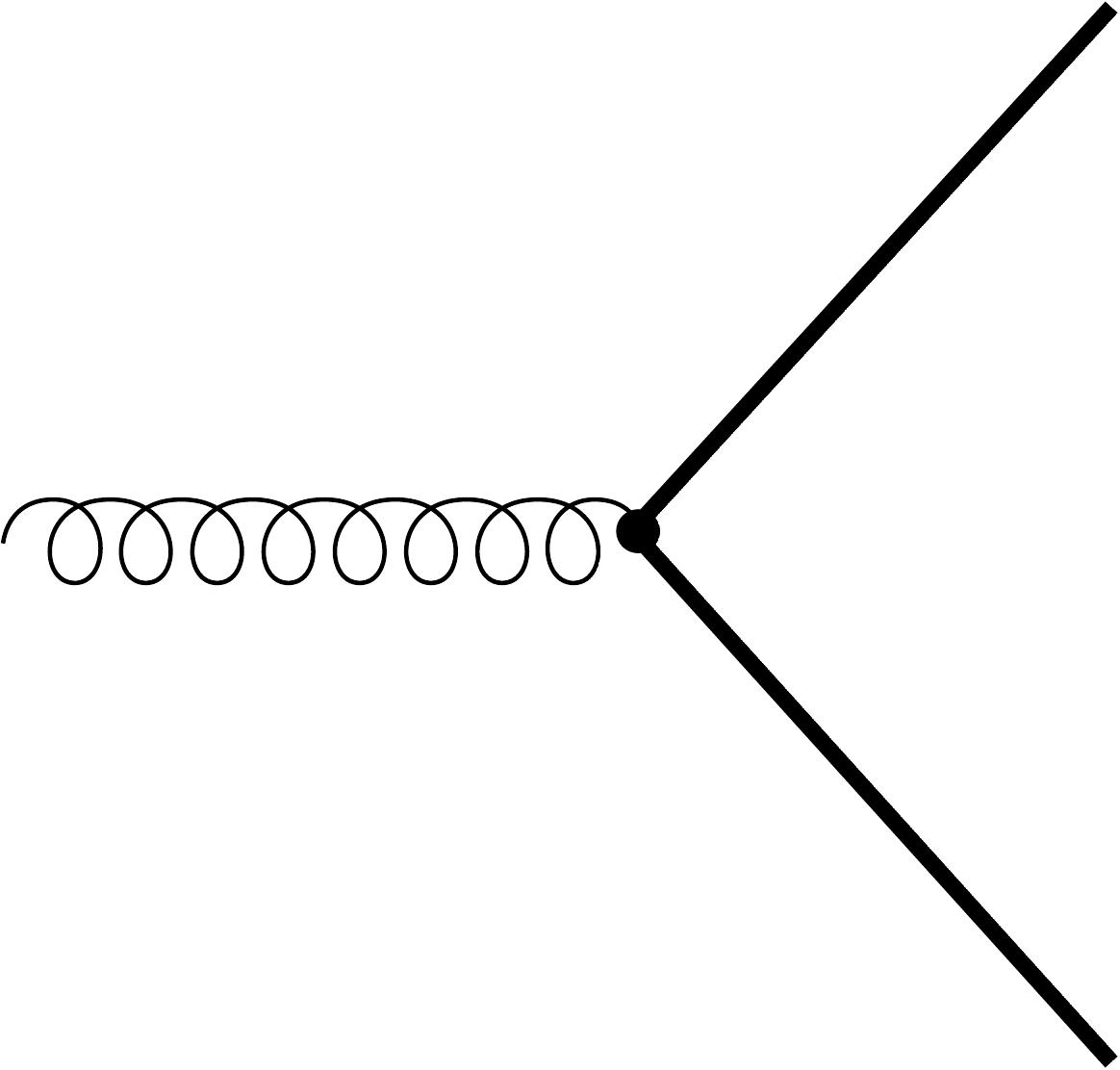}}}
\; .
\end{equation}
In fact, the vacuum polarization insertions could be Dyson-resummed
leading to the replacement
\begin{equation}
\Pi(s) \; \rightarrow \; \frac{1}{1-\Pi(s)} \; .
\end{equation}
By unitarity, the production cross section in the octet configuration
is proportional to $\mbox{Im}\,\Pi(s)$, which means that the result
containing annihilation contributions to all orders is given by
$\mbox{Im}\,1/(1-\Pi(s))$. Returning, however, to our fixed-oder
analysis at next-to-next-to-leading order, we note that we only have
to consider a single diagram, Fig.~\ref{fig:potential} b). The vacuum
polarization contribution in this diagram will be denoted by
$\Pi^{(2)}_C(s)$. As long as we are only interested in Coulomb
exchanges, the vacuum polarization close to threshold is expressed
through the non-relativistic Coulomb Green function
\cite{Penin:1998zh} with appropriate color-factor changes to take into
account the octet configuration. Nevertheless, we will need much less
information here. Indeed, if we consider the imaginary part of
$\Pi^{(2)}_C(s)$, we can write
\begin{equation}
\mbox{Im} \, \Pi^{(2)}_C(s) \approx \frac{\pi\as}{2\beta} \left( \cf -
\frac{\ca}{2} \right) \, \mbox{Im} \, \Pi^{(1)}(s) \approx
\frac{\pi\as}{2\beta} \left( \cf - \frac{\ca}{2} \right)
\left(- \frac{1}{2} \as \beta \, \tf \right) \; ,
\end{equation}
where, in the first approximation, we have used the fact that the
imaginary part of the vacuum polarization is related to the cross
section, which receives a Coulomb correction in the form of the
Sommerfeld factor, $\pi \alpha/\beta/(1-\exp(-\pi \alpha/\beta))$,
expanded to first non-trivial order, where now $\alpha = (\cf - \ca/2)
\as$. The second approximation follows from the well-known (textbook)
expression for the one-loop vacuum polarization function
$\Pi^{(1)}(s)$. The $\beta$ suppression of the imaginary part due to
the vanishing phase space volume at threshold is canceled by the
Sommerfeld $1/\beta$ enhancement. The expression thus tends to a
non-vanishing constant at threshold. This is the reason for which the
next-to-leading order cross section for top-quark pair-production also
tends to a non-vanishing constant.

At this point, we have only determined the imaginary part of the
vacuum polarization in a particular limit, which does not allow us to
use dispersion relations to obtain the real part. However, it is still
possible to use unitarity and analyticity for the same
purpose. Indeed, the singularity at $\beta = 0$ is a branch point
modeled as usual by a logarithm. An  appropriate variable to express
the occurrence of the cut is the non-relativistic energy of the system,
$E = \sqrt{s} - 2m \approx m \beta^2$, where the last approximation is
valid for $\beta \approx 0$. The imaginary part across the cut is
obtained from
\begin{equation}
\ln\left( \frac{-E - i \epsilon}{m} \right) \approx 2 \ln \beta - i\pi
\; .
\end{equation}
Thus, the fact that the imaginary part tends to a constant together
with the presence of a cut in the non-relativistic energy starting
from $E = 0$, allows us to write the final result for the contribution
to the cross section
\begin{equation}
\label{eq:VP}
\left( \frac{\as}{4\pi} \right)^2
\sigma^{(2,0)}_{q\bar{q},\bf{8}}\Big|_{\rm VP}
\approx 2 \, \mbox{Im} \, \Pi^{(2)}_C(s) \, \mbox{Re} \, \left(
-\frac{1}{\pi} \ln\left( \frac{-E - i \epsilon}{m} \right) \right)
\approx \left( \frac{\as}{4\pi} \right)^2 \left(\cf - \frac{\ca}{2}
\right) 8\pi^2 \ln \beta \; .
\end{equation}

Let us now apply the factorization formula
Eq.~(\ref{eq:factorization}). For the case at hand, we write
\begin{equation}
\label{eq:factorization2}
\hat \sigma_{ij, \, \alpha} = \hat \sigma^{(2), \, VV, \, fin}_{ij,
  \alpha} + \hat \sigma^{(1), \, V, \, fin}_{ij, \alpha} \otimes
S^{(1)}_{ij, \alpha} + \hat \sigma^{(0)}_{ij, \alpha} \otimes
S^{(2)}_{ij, \alpha} \; ,
\end{equation}
where $\hat \sigma^{(1), \, V, \, fin}_{ij,\alpha}$ and $\hat
\sigma^{(2), \, VV, \, fin}_{ij,\alpha}$ are cross sections evaluated
with the finite remainder of the virtual amplitudes at the next-to-
and next-to-next-to-leading orders. Notice that the latter require the
one-loop-squared corrections \cite{Korner:2008bn, Kniehl:2008fd,
  Anastasiou:2008vd}, besides the two-loop amplitudes of the present
publication. In order to obtain the $\beta$-expansion of the
one-loop-squared finite remainders, we have used our own results from
\cite{Czakon:2013goa}. The soft functions, $S^{(i)}_{ij,\alpha}$, on
the other hand, can be found in Ref.~\cite{Belitsky:1998tc} for the
singlet, and in Ref.~\cite{Czakon:2013hxa} for the octet
configurations of the final state. They are only affected by the
choice of the initial state through the Casimir invariants of the
respective representations of the gauge group.

Results for $\hat \sigma_{ij, \, \alpha}$ with exact dependence on the
number of colors and active flavors can be found in
Appendix~\ref{sec:expansions}. Here, we reproduce the results with
$N_c = 3$ and $\nl = 5$. For the gluon-fusion channel, there is
\cite{Beneke:2009ye}
\begin{eqnarray}
\sigma^{(2)}_{gg} &=&
\frac{68.5471}{\beta^2}
+\frac{1}{\beta}\Big(496.3\lb2+321.137\lb1-8.62261\Big) \nonumber \\ &&
+4608\lb4-1894.91\lb3-912.349\lb2+2456.74\lb1+C^{(2)}_{gg} \; ,
\end{eqnarray}
where the constant is obtained by combining the contributions from the
three color configurations
\begin{equation}
C^{(2)}_{gg} = \frac{C^{(2)}_{gg, \bf 1} \, \hat \sigma^{(0)}_{gg,\bf 1}
  + C^{(2)}_{gg, \bf 8_S} \, \hat \sigma^{(0)}_{gg,\bf 8_S}}{\hat
  \sigma^{(0)}_{gg,\bf 1} + \hat \sigma^{(0)}_{gg,\bf 8_S}} +
C^{(2)}_{gg, \bf 8_A} \; ,
\end{equation}
with
\begin{eqnarray}
C^{(2)}_{gg, \bf 1} &=& 37.1457 + 17.2725 \, \nl = 123.508 \; ,
\nonumber \\
C^{(2)}_{gg, \bf 8_S} &=& 674.517 - 45.5875 \, \nl = 446.58 \; ,
\nonumber \\
C^{(2)}_{gg, \bf 8_A} &=& 11.2531 - 2.29745 \, \nl + 0.142857 \, \nl^2
= 3.33731 \; ,
\end{eqnarray}
where the last number here and below corresponds to $\nl = 5$.
Combining these numbers leads to
\begin{equation}
C^{(2)}_{gg} = 503.664 - 29.9249 \, \nl + 0.142857 \, \nl^2 = 357.611 \; .
\end{equation}
This result should be compared to the one obtained by expanding the
fitting formulae for the total cross section presented in
\cite{Czakon:2013goa}
\begin{equation}
C^{(2), \, \mbox{\cite{Czakon:2013goa}}}_{gg} = 338.179 - 26.8912 \, \nl +
0.142848 \, \nl^2 = 207.294 \; .
\end{equation}
The correct value of the coefficient at $\nl = 0$ differs from that
from Ref.~\cite{Czakon:2013goa} by about 50\%, which fits within the error
estimate from the latter publication. A partial cancellation of this
coefficient by the term proportional to $\nl$ is responsible for the
even larger difference of the values at $\nl = 5$.

In the quark-annihilation channel we have
\begin{eqnarray}
\sigma^{(2)}_{q\bar{q}} &=&
\frac{3.60774}{\beta^2}
+\frac{1}{\beta}\Big(-140.368\lb2+32.106\lb1+3.95105\Big) \nonumber \\ &&
+910.222\lb4-1315.53\lb3+592.292\lb2+515.397\lb1+C^{(2)}_{q\bar q} \; .
\end{eqnarray}
This formula differs from the one presented in
Ref.~\cite{Beneke:2009ye}, in the coefficient in front of $\lb1$, by
the amount given in Eq.~(\ref{eq:VP}). In
the case of top-quarks, i.e. with $\nl=5$, the numerical effect of the
correction is tiny. It amounts to a 2.5\% reduction of the
coefficient, which was quoted to be 528.557 in
Ref.~\cite{Beneke:2009ye}. This is important, as the results for the
exact total cross sections in Ref.~\cite{Baernreuther:2012ws} have
been parameterized with respect to the threshold expansion. It turns
out that the difference in the coefficient of $\lb1$ is invisible
within the numerical precision of the results of
Ref.~\cite{Baernreuther:2012ws}. For the constant term, we obtain
\begin{equation}
C^{(2)}_{q\bar q} = 1104.08 - 42.9666 \, \nl - 4.28168 \, \nl^2 = 782.208 \; ,
\end{equation}
which should be compared with the value from the fitting formula of
Ref.~\cite{Baernreuther:2012ws}
\begin{equation}
C^{(2), \, \mbox{\cite{Baernreuther:2012ws}}}_{q\bar q} = 1195.82 - 44.1841 \, \nl
- 4.28168 \, \nl^2 = 867.858 \; .
\end{equation}
The two results are compatible at the 10\% level, which is exactly the
uncertainty quoted in Ref.~\cite{Baernreuther:2012ws}.


\section{Conclusions and outlook}

With this publication, we have completed the numerical analysis of
two-loop amplitudes for heavy-quark pair production in the 
quark-annihilation and gluon-fusion channels. We have demonstrated that
numerical methods based on differential equations lead to high
precision results in the relevant kinematical range. We have also
provided expansions in the threshold and high energy limits, which can
be used to obtain reliable numbers for any phase space points. These
results should also be viewed as benchmarks for future analytic
evaluations. In any case, they will constitute the basis for the
calculation of differential distributions for top-quark pair
production, which is the subject of current work.

Besides amplitudes, we were also able to provide threshold expansions
of partonic cross sections improving over previous studies. Our
results can be incorporated into resummed predictions for total cross
sections. Once virtual amplitudes become known completely
analytically, it will be possible to give fully analytic results for
the velocity independent terms of the threshold expansions. We have
provided formulae, which should make such an exercise straightforward.


\acknowledgments This research was supported by the German Research
Foundation (DFG) via the Sonderforschungsbereich/Transregio SFB/TR-9
``Computational Particle Physics''. The work of M.C. was supported by
the DFG Heisenberg programme.


\appendix

\section{Renormalization constants and anomalous dimensions}
\label{sec:RConst}

For convenience of the reader, we reproduce in this appendix the renormalization
and decoupling constants necessary for the renormalization of the
two-loop amplitudes. Notice that the one-loop contributions to the
constants have to be expanded up to ${\cal O}(\ep^2)$, since they will
multiply one-loop amplitudes, which contain soft-collinear divergences
leading to $1/\ep^2$ poles.

The on-shell renormalization constants are
\begin{eqnarray}
Z_g &=& 1
+ \left(\frac{\asnf}{2\pi}\right) \tf \nh
\biggl\{
  - {2 \over 3 \ep}
  - {2 \over 3} \lmu{}
  - {1 \over 3} \ep \lmu2
  - {\pi^2 \over 18} \ep
  - {1 \over 9} \ep^2 \lmu3
  - {\pi^2 \over 18} \ep^2 \lmu{}
  + {2 \over 9} \ep^2 \z3
\biggr\}
\nonumber\\
& &
+ \left(\frac{\asnf}{2\pi}\right)^2 \tf \nh
\biggl\{
  \tf \nh \biggl[
      {4 \over 9 \ep} \lmu{}
    + {2 \over 3} \lmu2
    + {\pi^2 \over 27}
  \biggr]
+ \tf \nl \biggl[
    - {4 \over 9 \ep^2}
    - {4 \over 9 \ep} \lmu{}
    - {2 \over 9} \lmu2
    - {\pi^2 \over 27}
  \biggr]
\nonumber\\
& &
+ \cf \biggl[
    - {1 \over 2 \ep}
    - \lmu{}
    - {15 \over 4}
  \biggr]
+ \ca \biggl[
      {35 \over 36 \ep^2}
    + {13 \over 18 \ep} \lmu{}
    - {5 \over 8 \ep}
    - {5 \over 4} \lmu{}
    + {1 \over 9} \lmu2
    + {13 \over 48}
    + {13 \pi^2 \over 216}
  \biggr]
\biggr\}\, ,
\nonumber \\ \nonumber \\
Z_q &=& 1 + \left(\frac{\asnf}{2\pi}\right)^2 \cf \tf \nh
\biggl[
  \frac{1}{4\ep}
  + \frac{1}{2} \lmu{}
  - \frac{5}{24}
\biggr]\, ,
\nonumber \\ \nonumber \\
Z_Q &=& 1
+ \left(\frac{\asnf}{2\pi}\right) \cf
\biggl\{
- \frac{3}{2\ep}
- 2
- \frac{3}{2}\lmu{}
- 4 \ep
- 2 \ep \lmu{}
- \frac{3}{4} \ep \lmu2
- \frac{\pi^2}{8} \ep
- 8 \ep^2
- 4 \ep^2 \lmu{}
- \ep^2 \lmu2
\nonumber\\
& &
- \frac{1}{4} \ep^2 \lmu3
- \frac{\pi^2}{6} \ep^2
- \frac{\pi^2}{8} \ep^2 \lmu{}
+ \frac{1}{2} \ep^2 \z3
\biggr\}
+ \left(\frac{\asnf}{2\pi}\right)^2 \cf
\biggl\{
\tf \nh \biggl[
\frac{1}{4\ep}
+ \frac{1}{\ep} \lmu{}
+ \frac{947}{72}
+ \frac{11}{6} \lmu{}
\nonumber\\
& &
+ \frac{3}{2} \lmu2
- \frac{5\pi^2}{4}
\biggr]
+ \tf \nl \biggl[
- \frac{1}{2\ep^2}
+ \frac{11}{12\ep}
+ \frac{113}{24}
+ \frac{19}{6} \lmu{}
+ \frac{1}{2} \lmu2
+ \frac{\pi^2}{3}
\biggr]
+ \cf \biggl[
\frac{9}{8\ep^2}
+ \frac{51}{16\ep}
\nonumber\\
& &
+ \frac{9}{4\ep} \lmu{}
+ \frac{433}{32}
+ \frac{51}{8} \lmu{}
+ \frac{9}{4} \lmu2
- \frac{49\pi^2}{16}
+ 4 \lt1 \pi^2
- 6 \z3
\biggr]
+ \ca \biggl[
\frac{11}{8\ep^2}
- \frac{127}{48\ep}
- \frac{1705}{96}
\nonumber\\
& &
- \frac{215}{24} \lmu{}
- \frac{11}{8} \lmu2
+ \frac{5\pi^2}{4}
- 2 \lt1 \pi^2
+3 \z3
\biggr] 
\biggr\}
\, ,
\nonumber \\ \nonumber \\
Z_m &=& 1
+ \left(\frac{\asnf}{2\pi}\right) \cf
\biggl\{
- \frac{3}{2\ep}
- 2
- \frac{3}{2}\lmu{}
- 4 \ep
- 2 \ep \lmu{}
- \frac{3}{4} \ep \lmu2
- \frac{\pi^2}{8} \ep
- 8 \ep^2
- 4 \ep^2 \lmu{}
- \ep^2 \lmu2
\nonumber\\
& &
- \frac{1}{4} \ep^2 \lmu3
- \frac{\pi^2}{6} \ep^2
- \frac{\pi^2}{8} \ep^2 \lmu{}
+ \frac{1}{2} \ep^2 \z3
\biggr\}
+ \left(\frac{\asnf}{2\pi}\right)^2 \cf
\biggl\{
\tf \nh \biggl[
- \frac{1}{2\ep^2}
+ \frac{5}{12\ep}
+ \frac{143}{24}
\nonumber\\
& &
+ \frac{13}{6} \lmu{}
+ \frac{1}{2} \lmu2
- \frac{2\pi^2}{3}
\biggr]
+ \tf \nl \biggl[
- \frac{1}{2\ep^2}
+ \frac{5}{12\ep}
+ \frac{71}{24}
+ \frac{13}{6} \lmu{}
+ \frac{1}{2} \lmu2
+ \frac{\pi^2}{3}
\biggr]
\nonumber\\
& &
+ \cf \biggl[
\frac{9}{8\ep^2}
+ \frac{45}{16\ep}
+ \frac{9}{4\ep} \lmu{}
+ \frac{199}{32}
+ \frac{45}{8} \lmu{}
+ \frac{9}{4} \lmu2
- \frac{17\pi^2}{16}
+ 2 \lt1 \pi^2
- 3 \z3
\biggr]
\nonumber\\
& &
+ \ca \biggl[
\frac{11}{8\ep^2}
- \frac{97}{48\ep}
- \frac{1111}{96}
- \frac{185}{24} \lmu{}
- \frac{11}{8} \lmu2
+ \frac{\pi^2}{3}
- \lt1 \pi^2
+ \frac{3}{2} \z3
\biggr] 
\biggr\}
\, ,
\end{eqnarray}
where $\lmu{} = \ln \mu^2/m^2$, and the first two formulae (on-shell
wave-function renormalization constants for the gluon and light quark
fields) have been taken from \cite{Czakon:2007ej, Czakon:2007wk},
while the third and fourth (heavy-quark wave-function and mass
renormalization constants) from \cite{Broadhurst:1991fy}.

The $\MSbar$ renormalization constant for the strong coupling up
to the two-loop level is given in terms of beta-function coefficients
\begin{eqnarray}
Z_\as =
   1
   - \left( {\asnf \over 2 \pi} \right) \frac{b_0}{2 \epsilon} 
   +  \left( {\asnf \over 2 \pi} \right)^2 \left(
     \frac{b_0^2}{4\epsilon^2}
     - \frac{b_1}{8 \epsilon}
     \right)
\, ,
\end{eqnarray}
where
\begin{eqnarray}
b_0 = {11 \over 3} \ca - {4 \over 3} \tf \nf \, ,
\qquad
b_1 = {34 \over 3} \ca^2 - {20 \over 3} \ca \tf \nf - 4 \cf \tf \nf \, .
\end{eqnarray}

Finally, we reproduce the two-loop decoupling constant for the strong coupling
\cite{Bernreuther:1981sg}
\begin{eqnarray}
\zeta_\as &=& 1
+ \left(\frac{\asnl}{2\pi}\right) \tf \nh
\biggl\{
\frac{2}{3} \lmu{}
+ \frac{1}{3} \ep \lmu2
+ \frac{\pi^2}{18} \ep
+ \frac{1}{9} \ep^2 \lmu3
+ \frac{\pi^2}{18} \ep^2 \lmu{}
- \frac{2}{9} \ep^2 \z3
\biggr\}
\nonumber\\
& &
+ \left(\frac{\asnl}{2\pi}\right)^2 \tf \nh
\biggl\{
\frac{4}{9} \tf \nh
\lmu2  
+ \cf \biggl[
\frac{15}{4}
+ \lmu{}
\biggr]
+ \ca \biggl[
- \frac{8}{9}
+ \frac{5}{3} \lmu{}
\biggr]
\biggr\}\, .
\end{eqnarray}

We also list the anomalous dimensions occurring in
Eq.~(\ref{eq:IRGamma}) necessary to obtain the finite remainders of
the two-loop amplitudes. The anomalous dimensions related to a single
parton (collinear in origin for massless partons and soft in origin
for massive partons) are \cite{Becher:2009qa, Becher:2009kw}
\begin{eqnarray}
\gamma^g\left( \asnl \right) &=& \left( \frac{\asnl}{2\pi} \right) 
\biggl\{
- \frac{11}{6} \ca
+ \frac{2}{3} \tf \nl
\biggr\}
+ \left( \frac{\asnl}{2\pi} \right)^2
\biggl\{
\ca^2 \biggr[
- \frac{173}{27}
+ \frac{11\pi^2}{72}
+ \frac{1}{2} \z3
\biggr]
\nonumber \\
& & 
+ \ca \tf \nl \biggl[
\frac{64}{27}
- \frac{\pi^2}{18}
\biggr]
+ \cf \tf \nl
\biggr\} \, ,
\\ \nonumber \\
\gamma^q\left( \asnl \right) &=& - \left( \frac{\asnl}{2\pi} \right)
\frac{3}{2} \cf
+ \left( \frac{\asnl}{2\pi} \right)^2 \cf
\biggl\{
\ca \biggr[
- \frac{961}{216}
- \frac{11\pi^2}{24}
+ \frac{13}{2} \z3
\biggr]
\nonumber \\
& & 
+ \cf \biggl[
- \frac{3}{8}
+ \frac{\pi^2}{2}
- 6 \z3
\biggr]
+ \tf \nl \biggl[
\frac{65}{54}
+ \frac{\pi^2}{6}
\biggr]
\biggr\} \, ,
\\ \nonumber \\
\gamma^Q\left( \asnl \right) &=& - \left( \frac{\asnl}{2\pi} \right) \cf
+ \left( \frac{\asnl}{2\pi} \right)^2 \cf
\biggl\{
\ca \biggr[
- \frac{49}{18}
+ \frac{\pi^2}{6}
- \z3
\biggr]
+ \frac{10}{9} \tf \nl
\biggr\} \, .
\end{eqnarray}

The cusp anomalous dimensions are given by \cite{Korchemsky:1987wg,
  Kidonakis:2009ev}
\begin{eqnarray}
\gamma_{\rm cusp}\left( \asnl \right) &=& \frac{\asnl}{\pi}
+\left( \frac{\asnl}{2\pi} \right)^2
\biggl\{
\ca \biggl[
\frac{67}{9}
- \frac{\pi^2}{3}
\biggr]
- \frac{20}{9} \tf \nl
\biggr\} \, ,
\\ \nonumber \\
\gamma_{\rm cusp}\left( \beta, \asnl \right) &=&
\gamma_{\rm cusp}\left( \asnl \right) \beta \coth \beta
\nonumber \\
& + &
\left( \frac{\asnl}{2\pi} \right)^2 2 \ca
\biggl\{
\coth^2\beta
\biggl[
{\rm Li}_3(e^{-2\beta})
+ \beta \, {\rm Li}_2(e^{-2\beta})
- \z3
+ \frac{\pi^2}{6} \beta
+ \frac{1}{3} \beta^3
\biggr]
\nonumber \\
& &
+ \coth\beta
\biggl[
{\rm Li}_2(e^{-2\beta})
-2\beta \ln (1 - e^{-2\beta})
- \frac{\pi^2}{6}(1 + \beta)
- \beta^2
- \frac{1}{3} \beta^3
\biggr]
\nonumber \\
& &
+ \frac{\pi^2}{6}
+ \z3
+ \beta^2
\biggr\} \, .
\end{eqnarray}
%


\section{Color dependence of threshold expansions}
\label{sec:expansions}

In this appendix, we present the threshold expansions of the partonic
cross sections with the normalization defined in
Eq.~(\ref{eq:result-scales}), including exact dependence on $N_c$, the
number of colors of the $\mbox{SU}(N_c)$ gauge group, and the number
of active flavours, $\nl$, at $\mu = m$. We decompose the results into
singular terms in $\beta$ and $\beta$-indpendent terms. First, we
reproduce the former from Ref.~\cite{Beneke:2009ye} after correcting
the quark-annihilation channel result as derived in
Section~\ref{sec:ThCross}
\begin{eqnarray}
\sigma^{(2,0)}_{q\bar{q}} &=&
\frac{(2\cf-\ca)^2\pi^4}{3\beta^2}+\frac{(2\cf-\ca)\pi^2}{9\beta}
\Big[288\cf\lb2+6\big(48\cf\lt1-23\ca+2\nl\big)\lb1 \nonumber \\ &&
+12\cf\big(-24+9\lt1+\pi^2\big)+3\ca\big(89-58\lt1-3\pi^2\big)+6\nl\big(-5
+6\lt1\big)-32\Big] \nonumber \\ &&
+512C_F^2\lb4+\frac{128}{9}\cf\Big[72\cf\big(-2+3\lt1\big)-29\ca+2\nl\Big]\lb3
\nonumber \\ &&
+\frac{16}{9}\Big[2\cf\big(12\cf(120-207\lt1+156\lt2-7\pi^2)+3\ca(217-198\lt1
-4\pi^2) \nonumber \\ &&
+6\nl(-9+10\lt1)-32\big)+3\ca(17\ca-2\nl)\Big]\lb2 \nonumber \\ &&
+\frac{4}{27}\Big[2\cf\big(36\cf(-960+\lt1(1368-84\pi^2)-1140\lt2+576\lt3
+55\pi^2+336\zeta_3) \nonumber \\[0.1cm] &&
+2\ca(-7582+108\lt1(115-2\pi^2)-5886\lt2+360\pi^2+189\zeta_3)
\nonumber \\[0.2cm] && +4\nl(338-630\lt1+378\lt2-9\pi^2)+768-1152\lt1+27\pi^2\big)
\nonumber \\[0.1cm] &&
+3\ca\big(6\ca(-185+126\lt1+6\pi^2-6\zeta_3)+12\nl(11-10\lt1)+64-9\pi^2\big)\Big]\lb1
\nonumber \\[0.2cm] && + C^{(2)}_{q\bar q} \; , \label{eq:sigma-2-0-qq}
\\[0.5cm] \nonumber 
\sigma^{(2,0)}_{gg,\bf 1} &=& \frac{4C_F^2
\pi^4}{3\beta^2}+\frac{2\cf\pi^2}{9\beta}\Big[288\ca\lb2
+6\big(\ca(-11+48\lt1)+2\nl\big)\lb1 \nonumber \\ &&
+9\cf\big(-20+\pi^2\big)+\ca\big(67-66\lt1+3\pi^2\big)+2\nl\big(-5
+6\lt1\big)\Big]+512C_A^2\lb4 \nonumber \\ &&
+\frac{128}{9}\ca\Big[\ca\big(-155+216\lt1\big)+2\nl\Big]\lb3
+\frac{32}{9}\ca\Big[9\cf\big(-20+\pi^2\big) \nonumber \\ &&
+\ca\big(1963-2790\lt1+1872\lt2-96\pi^2\big)+2\nl\big(-17+18\lt1\big)\Big]\lb2
\nonumber \\ &&
+\frac{16}{27}\Big[27\cf\big(-2\cf\pi^2+\ca(80+6\lt1(-20+\pi^2)-5\pi^2)\big)
+\ca\big(\ca(-23758 \nonumber \\[0.2cm] &&
+18\lt1(1963-96\pi^2)-24246\lt2+10368\lt3+1251\pi^2+6237\zeta_3)
\nonumber \\[0.2cm] && +2\nl(218-306\lt1+162\lt2-9\pi^2)\big)\Big]\lb1
+ C^{(2)}_{gg,\bf 1} \; , \label{eq:sigma-2-0-gg1}
\\[0.5cm] \nonumber 
\sigma^{(2,0)}_{gg,\bf 8_S} &=&
\frac{(2\cf-\ca)^2\pi^4}{3\beta^2}+\frac{(2\cf-\ca)\pi^2}{18\beta}
\Big[576\ca\lb2+12\big(\ca(-23+48\lt1)+2\nl\big)\lb1 \nonumber \\ &&
+18\cf\big(-20+\pi^2\big)+\ca\big(278-132\lt1-3\pi^2\big)+4\nl\big(-5
+6\lt1\big)\Big] +512C_A^2\lb4 \nonumber \\ &&
+\frac{128}{9}\ca\Big[\ca\big(-173+216\lt1\big)+2\nl\Big]\lb3+\frac{16}{9}\ca
\Big[18\cf\big(-20+\pi^2\big) \nonumber \\ &&
+\ca\big(4553-6156\lt1+3744\lt2-201\pi^2\big)+2\nl\big(-37+36\lt1\big)\Big]\lb2
\nonumber \\ &&
+\frac{4}{27}\Big[54\cf\big(-4\cf\pi^2+\ca(180+12\lt1(-20+\pi^2)-7\pi^2)\big)
+\ca\big(\ca(-111418 \nonumber \\[0.2cm] &&
+36\lt1(4499-201\pi^2)-105624\lt2+41472\lt3+5823\pi^2+24840\zeta_3)
\nonumber \\[0.2cm] && +4\nl(505-666\lt1+324\lt2-18\pi^2)\big)\Big]\lb1
 + C^{(2)}_{gg,\bf 8_S} \; . \label{eq:sigma-2-0-gg8}
\end{eqnarray}
The $\beta$-independent terms are obtained from
Eq.~(\ref{eq:factorization2}). They read
\begin{equation}
\begin{split}
C^{(2)}_{q\overline{q}}&=N_c^2 \left(\frac{146218}{27}-\frac{33677 \pi ^2}{108}-\frac{491 \pi ^4}{240}-\frac{11773 \zeta (3)}{9}+\left(-8044+\frac{815 \pi ^2}{2}+1404 \zeta (3)\right) \lt1 \right.\\
              &\left. +\left(\frac{50585}{9}-209 \pi ^2\right) \lt2 -2252 \lt3 +508 \lt4 \right)\\
              & + N_c \left(-\frac{13568}{81}+\frac{56 \pi ^2}{9}+\frac{1952}{9}\lt1 -\frac{1088}{9}\lt2\right)\\
              & + \left(-\frac{588430}{81}+\frac{43709 \pi ^2}{108}+\frac{467 \pi ^4}{72}+\frac{19891 \zeta (3)}{9}+\left(\frac{97336}{9}-604 \pi ^2-2772 \zeta (3)\right) \lt1 \right.\\
              & \left. +\left(-\frac{70286}{9}+372 \pi ^2\right) \lt2 +3584 \lt3-1016 \lt4\right)\\
              & + \frac{1}{N_c}\left(\frac{256}{3}-\frac{40 \pi ^2}{9}+\left(-160+2 \pi ^2\right) \lt1 +\frac{1088}{9}\lt2\right)\\
              & + \frac{1}{N_c^2}\left(2624-\frac{1525 \pi ^2}{12}-\frac{533 \pi ^4}{144}-896 \zeta (3)+\left(-3696+\frac{1253 \pi ^2}{6}+1344 \zeta (3)\right) \lt1\right.\\
              & \left.  +\left(2505-\frac{472 \pi ^2}{3}\right) \lt2 -1332 \lt3+508 \lt4 \right)\\
              & + n_l \left(N_c\left(-\frac{23380}{81}+\frac{685 \pi \
^2}{54}+\frac{226 \zeta (3)}{9}+\left(\frac{13940}{27}-\frac{38 \pi ^2}{3}\right) \lt1-\frac{3584}{9}\lt2\right.\right.\\
              &\left. \left. +\frac{416}{3}\lt3 \right)+ \left(\frac{320}{81}-\frac{128}{27}\lt1 \right) + \frac{1}{N_c}\left(\frac{14948}{81}-\frac{661 \pi^2}{54}-\frac{226 \zeta (3)}{9}\right. \right.\\
              & \left. \left. +\left(-\frac{3220}{9}+14 \pi^2\right) \lt1 + \frac{2984}{9} \lt2 -\frac{416}{3} \lt3 \right)\right)\\
              & + n_l^2 \left(\frac{100}{81}+\frac{4 \pi
  ^2}{9}-\frac{80}{27} \lt1 +\frac{16}{9} \lt2\right)\\
& + 2 \int d \cos\theta \; \frac{2 \, \mbox{Re} \, \langle {\cal M}_q^{(0)} | {\cal
  M}_q^{(2), \, fin} \rangle}{\langle {\cal M}_q^{(0)} | {\cal
    M}_q^{(0)} \rangle} \Biggl|_{\beta^0} \; ,
\end{split}
\end{equation}
\begin{equation}
\begin{split}
C^{(2)}_{gg,{\bf 1}}&=N_c^2 \left(\frac{1091701}{81}-\frac{19414 \pi ^2}{27}-\frac{1895 \pi ^4}{144}-\frac{35750 \zeta (3)}{9}\right.\\
                    & \left. +\left(-\frac{181424}{9}+\frac{3181 \pi ^2}{3}+5544 \zeta (3)\right) \lt1 +\left(14496-690 \pi ^2\right) \lt2 -6672 \lt3\right.\\
                    & \left.  +2032 \lt4   \right) + \left(610-\frac{188 \pi ^2}{3}-\frac{5 \pi ^4}{8}+\left(-960+\frac{166 \pi ^2}{3}\right) \lt1 +\left(680-42 \pi ^2\right) \lt2 \right)\\
                    & + \frac{1}{N_c^2}\left(25-\frac{\pi ^2}{6}+\frac{17 \pi ^4}{16}-\frac{11}{3} \pi ^2 \lt1+4 \pi ^2 \lt2  \right)\\
                    & + n_l\,N_c \left(-\frac{21256}{81}+\frac{403 \pi ^2}{27}+\frac{452 \zeta (3)}{9}+\left(\frac{3488}{9}-16 \pi ^2\right) \lt1-272 \lt2 +96 \lt3   \right)\\
& + 2 \int d \cos\theta \; \frac{2 \, \mbox{Re} \, \langle {\cal
    M}_g^{(0)} | {\cal P}_{\bf 1}| {\cal M}_g^{(2), \, fin}
  \rangle}{\langle {\cal M}_g^{(0)} |{\cal P}_{\bf 1}| {\cal
    M}_g^{(0)} \rangle} \Biggl|_{\beta^0} \; ,
\end{split}
\end{equation}
\begin{equation}
\begin{split}
C^{(2)}_{gg,{\bf 8_S}}&=N_c^2 \left(\frac{1285909}{81}-\frac{46937 \pi^2}{54}-\frac{253 \pi ^4}{20} -\frac{39728 \zeta (3)}{9}\right.\\
                       & \left. +\left(-\frac{210416}{9}+1228 \pi^2+5520 \zeta (3)\right) \lt1 +\left(16464-740 \pi^2\right) \lt2 -7168 \lt3\right.\\
                       & \left. +2032 \lt4   \right) + \left(750-\frac{135 \pi ^2}{2}+\frac{3 \pi ^4}{2}+\left(-1060+53 \pi ^2\right) \lt1 +\left(680-34 \pi ^2\right) \lt2 \right)\\
                       & + \frac{1}{N_c^2}\left(25-\frac{\pi ^2}{6}+\frac{17 \pi ^4}{16}-\frac{11}{3} \pi ^2 \lt1 +4 \pi ^2 \lt2 \right)\\
                       &  + n_l\,N_c \left(-\frac{25528}{81}+\frac{439 \pi ^2}{27}+\frac{452 \zeta (3)}{9}+\left(\frac{4040}{9}-16 \pi ^2\right) \lt1-296 \lt2  +96 \lt3  \right)\\
& + 2 \int d \cos\theta \; \frac{2 \, \mbox{Re} \, \langle {\cal
    M}_g^{(0)} |{\cal P}_{\bf 8_S}| {\cal M}_g^{(2), \, fin}
  \rangle}{\langle {\cal M}_g^{(0)} | {\cal P}_{\bf 8_S}|{\cal
    M}_g^{(0)} \rangle} \Biggl|_{\beta^0} \; .
\end{split}
\end{equation}
The remaining integrals must be performed with the numerical values
given in Tabs.~\ref{tab:constgg} and \ref{tab:constqq}. The
restriction $|_{\beta^0}$ is there to specify that only the
$\beta$-independent contributions of the amplitude expansions are to
be included.

The $\beta$-independent term in the case of the anti-symmetric octet
configuration in the gluon-fusion channel, which is also the leading term in
the expansion for this color configuration, can be given in an
entirely analytic form
\begin{equation}
\begin{split}
C^{(2)}_{gg,{\bf 8_A}}&=\frac{1}{144(N_c^2-2)}\left(N_c^4 \left(64-384 \lt1+576 \lt2  \right) + N_c^3 \left(448-48 \pi ^2\right.\right.\\
                      & \left. \left. +\left(-1344+144 \pi ^2\right) \lt1\right) + N_c^2 \left(976-264 \pi ^2+9 \pi ^4+\left(192+288 \pi ^2\right) \lt1\right.\right.\\
                      & \left. \left. -2304 \lt2 \right) + N_c \left(672-408 \pi ^2+36 \pi ^4+\left(2688-288 \pi ^2\right) \lt1 \right) + \left(144-96 \pi ^2\right.\right.\\
                      & \left. \left. +52 \pi ^4+\left(1152-576 \pi ^2\right) \lt1 +2304 \lt2  \right) + n_l \left(N_c^3 \left(64-192 \lt1\right) \right.\right.\\
                      & \left. \left. + N_c^2 \left(224-24 \pi
^2\right) + N_c \left(96-48 \pi ^2+384 \lt1\right)\right)+
16\,n_l^2\,N_c^2\right) \; .
\end{split}
\end{equation}


\end{document}